\documentclass[sigconf]{acmart}

\AtBeginDocument{%
  }

\copyrightyear{2025}
\acmYear{2025}
\setcopyright{cc}
\setcctype{by}
\acmConference[CHI '25]{CHI Conference on Human Factors in Computing Systems}{April 26-May 1, 2025}{Yokohama, Japan}
\acmBooktitle{CHI Conference on Human Factors in Computing Systems (CHI '25), April 26-May 1, 2025, Yokohama, Japan}\acmDOI{10.1145/3706598.3713913}
\acmISBN{979-8-4007-1394-1/25/04}

\usepackage{enumitem}
\usepackage{xargs}      % Used for new commands with optional arguments
\usepackage{soul}       % Used for custom comments
\usepackage{color}      % Used for custom colors in comments
\usepackage{xspace}     % Used for abbreviation spacing
\usepackage{xpunctuate} % Used for abbreviation spacing
\usepackage{multirow}
\usepackage{array}
\usepackage{multirow}
\usepackage{tabularx}
\usepackage{listings}
\usepackage{mdframed}
\usepackage{caption}

\newcommand{\sys}{\textsc{Jupybara}}

\newcommand{\ie}{{i.e.,}\xspace}
\newcommand{\eg}{{e.g.,}\xspace}
\newcommand{\etal}{{et~al\xperiod}\xspace}

\definecolor{lightpink}{RGB}{237,157,202}
\definecolor{lightred}{RGB}{210,121,121}
\definecolor{lightorange}{RGB}{230,170,50}
\definecolor{lightgold}{RGB}{210,194,121}
\definecolor{lightgreen}{RGB}{121,210,121}
\definecolor{lightaqua}{RGB}{121,206,210}
\definecolor{lightblue}{RGB}{121,124,210}
\definecolor{lightpurple}{RGB}{153,102,255}
\definecolor{red}{RGB}{178,34,34}
\definecolor{gray}{RGB}{166,166,166}
\definecolor{forestgreen}{RGB}{74,103,65}
\definecolor{DeepTeal}{HTML}{00796B}
\definecolor{RoyalBlue}{HTML}{4169E1}
\definecolor{BurntSienna}{HTML}{E97451}

\newcommand{\semanticcolor}{\textcolor{DeepTeal}}
\newcommand{\rhetoriccolor}{\textcolor{RoyalBlue}}
\newcommand{\pragmaticcolor}{\textcolor{BurntSienna}}

\newcommand{\cut}[1]{\textcolor{red}{}}

\usepackage{soul,color}
\soulregister\cite7
\soulregister\ref7
\soulregister\pageref7

\renewcommand{\hl}[1]{#1}

\begin{document}

\title{\sys: Operationalizing a Design Space for Actionable Data Analysis and Storytelling with LLMs}


\author{Huichen Will Wang}
\email{wwill@cs.washington.edu}
\affiliation{
  \institution{University of Washington\\ Tableau Research}
  \city{Seattle}
  \state{WA}
  \country{USA}
}

\author{Larry Birnbaum}
\email{l-birnbaum@northwestern.edu}
\affiliation{%
  \institution{Northwestern University}
  \city{Evanston, IL}
  \country{USA}
}

\author{Vidya Setlur}
\email{vsetlur@tableau.com}
\affiliation{%
  \institution{Tableau Research}
  \city{Palo Alto, CA}
  \country{USA}
}

\renewcommand{\shortauthors}{Wang et al.}

\begin{abstract}
 % Need to be max 150 words
\begin{figure*}[h!]
  \centering
  \includegraphics[width=\textwidth]{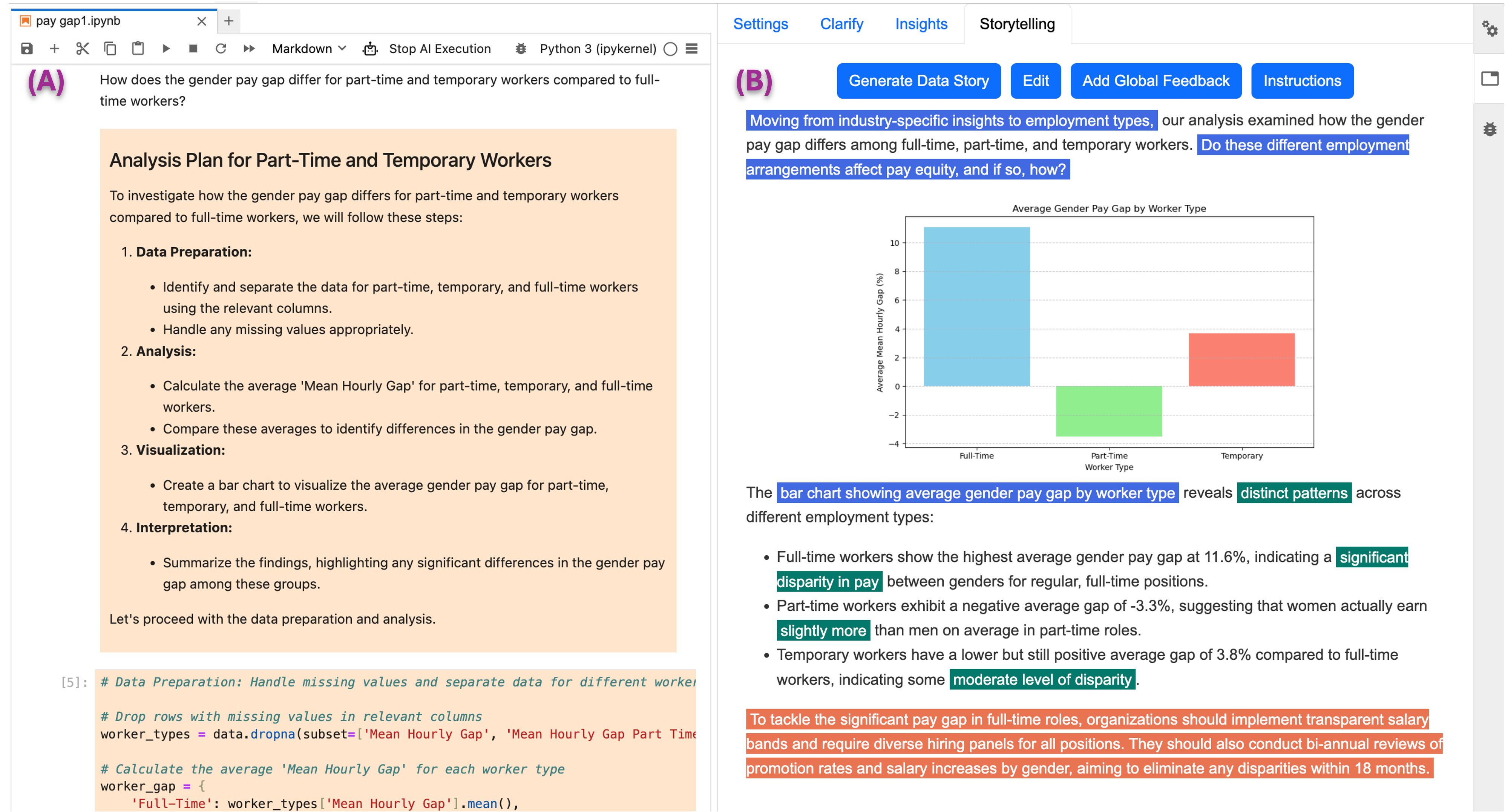}
  \caption{The interface of \sys, an AI-enabled assistant for actionable EDA and data storytelling implemented as a Jupyter Notebook extension. \sys\ operationalizes our proposed design space consisting of the \semanticcolor{semantic}, \rhetoriccolor{rhetorical}, and \pragmaticcolor{pragmatic} dimensions. (A) For a complex user query in EDA, \sys\ \rhetoriccolor{identifies and presents an analysis plan} before producing code. (B) In the data story generated by \sys, the system uses \semanticcolor{precise language} to convey analytical results; \rhetoriccolor{appropriate hooks, connectives, and narration of analytical strategies} to bolster actionable insights; and \pragmaticcolor{relevant domain knowledge} to connect data facts to actionable insights.}
  \label{fig:teaser}
\end{figure*}

Mining and conveying actionable insights from complex data is a key challenge of exploratory data analysis (EDA) and storytelling. To address this challenge, we present a design space for actionable EDA and storytelling. Synthesizing theory and expert interviews, we highlight how \semanticcolor{semantic precision}, \rhetoriccolor{rhetorical persuasion}, and \pragmaticcolor{pragmatic relevance} underpin effective EDA and storytelling. We also show how this design space subsumes common challenges in actionable EDA and storytelling, such as identifying appropriate analytical strategies and leveraging relevant domain knowledge. Building on the potential of LLMs to generate coherent narratives with commonsense reasoning, we contribute \sys, an AI-enabled assistant for actionable EDA and storytelling implemented as a Jupyter Notebook extension. \sys\ employs two strategies---design-space-aware prompting and multi-agent architectures---to operationalize our design space. An expert evaluation confirms \sys's usability, steerability, explainability, and reparability, as well as the effectiveness of our strategies in operationalizing the design space framework with LLMs.

\end{abstract}






\maketitle

% interface, co-design screenshot, architecture, user scenario flow

\section{Introduction}

\hl{Ben Schneiderman famously noted, ``the purpose of visualization is insight, not pictures.''~\cite{benquote}} Similarly, the goals of data analysis and storytelling extend beyond merely generating statistical output, data visualizations, or even data narratives: these activities are fundamentally about extracting and communicating insights.  

\hl{In this work,} we focus on a particular class of insights, \textit{actionable} insights. \hl{Traditionally, researchers have viewed insights as a collection of knowledge over the data~\cite{battle2023we, chang2009defining}, be it data facts, utterances, or knowledge links. A defining hallmark of actionable insights is that they not only} \semanticcolor{sit on knowledge over the data} \hl{but also aim to} \pragmaticcolor{empower people to make informed decisions and take sensible actions,} \hl{thus often} \rhetoriccolor{requiring strategies tailored to real-world questions} \hl{to derive}. \hl{The decision- and action-oriented nature of these insights makes them challenging to generate but crucial for solving practical problems.} In practice, many analytic processes are directed toward generating actionable insights. By combining data analytics with domain knowledge, these processes bridge the gap between raw data and strategic actions. For example, the entire practice of business analytics centers on transforming data into actionable insights that drive competitive advantage and organizational success by leveraging industry know-how~\cite{delen2018research}. Educational data mining synthesizes educational psychology with data mining techniques to resolve challenges in learning~\cite{romero2010educational}. An important branch of epidemiology is concerned with analyzing data to propose interventions for the spread of diseases that are grounded in biological and clinical theories~\cite{brachman1996epidemiology}. 

In this work, we interviewed nine expert data analysts to map a design space for exploratory data analysis and storytelling for actionable insights (hereafter actionable EDA and storytelling\footnote{\hl{We define actionable EDA as the analytical process of deriving actionable insights, and actionable data storytelling as the process of crafting a data story to convey actionable insights to an audience.}}) \hl{and considered them in foundational theories in data science, narrative discourse, and communication}. We illustrate how \semanticcolor{semantic precision} (accurately interpreting, tracking, and verbalizing EDA findings), \rhetoriccolor{rhetorical persuasion} (strategically mining and communicating results to corroborate actionable insights), and \pragmaticcolor{pragmatic relevance} (integrating EDA findings with domain expertise) collectively contribute to sound and effective actionable EDA and storytelling. The interviews also helped us characterize the workflows of actionable EDA and storytelling (fluid, integrated, and iterative) and uncover key challenges in the process, such as identifying appropriate analytical strategies and leveraging relevant domain knowledge.

We then show how our proposed design space can inform the development of better AI assistants for actionable EDA and storytelling. While these tasks have traditionally relied on human expertise, Large Language Models (LLMs) present exciting opportunities to augment human expertise with machine intelligence. With their statistical knowledge, LLMs can amplify human capabilities in extracting meaningful patterns from data. Furthermore, trained on vast corpora, LLMs can potentially generate coherent narratives and apply domain expertise to interpret analytical results, enabling analysts to make more informed strategic decisions. We propose two strategies to utilize LLMs in operationalizing the framework: design-space-aware prompting (prompting LLMs with guidance distilled from the design space) and multi-agent architectures (improving response quality along the dimensions \hl{of the design space} through multi-agent collaboration). We further implement these strategies in a Jupyter Notebook plugin supporting actionable EDA and storytelling, \sys\footnote{``\textsc{Jupybara}'' is a portmanteau of ``Jupyter'' (Notebook) and ``capybara,'' a friendly and likable creature that satisfies Gricean Maxims of cooperative behavior. Known for their social nature and tendency to stack on top of each other, capybaras symbolize the multi-agent architecture of the tool for facilitating actionable EDA and storytelling.} (Figure~\ref{fig:teaser}). In addition to being an agentic EDA copilot and data storyteller, \sys\ allows users to seek clarification, track insights, and edit AI-generated content manually or with AI assistance.

A preliminary evaluation with nine data analysts showed that participants found \sys\ produced higher-quality responses for actionable EDA and storytelling compared to web-based LLM tools (\eg ChatGPT). Participants also rated \sys\ as more usable, steerable, explainable, and reparable. Moreover, participants deemed the multi-agent mode of \sys\ to produce better responses than the single-agent mode across the three dimensions of our design space. In sum, our primary contributions are:

\begin{itemize}
    \item A design space that integrates the \semanticcolor{semantic}, \rhetoriccolor{rhetorical}, and \pragmaticcolor{pragmatic} dimensions to facilitate effective actionable EDA and storytelling;
    \item A formative interview and \hl{participatory design session} with nine data analysts that identified analysts' workflows, considerations, and challenges in actionable EDA and storytelling;
    \item Two strategies to operationalize the design space, design-space-aware prompting and multi-agent architectures, which we implement in a Jupyter Notebook plugin supporting actionable EDA and storytelling, \sys;
    \item A user evaluation of \sys\ with nine participants demonstrating the usability, steerability, explainability, and reparability of \sys\ and the effectiveness of our strategies for operationalizing the design space.
   
\end{itemize}

\section{Related Work}

Our work is informed by research in the use of language in visual analytics and human-AI collaboration for data analysis and storytelling.

\subsection{The Use of Language in Visual Analytics}
The integration of natural language processing (NLP) techniques with visual analytics tools has explored how language can be used to express user intent, bridging analytical inquiry and visualization responses. For example, several systems have proposed techniques for enabling users to query and explore data through conversational interfaces, addressing challenges related to linguistic ambiguity in user utterances~\cite{datatone,eviza,analyza}, as well as the conversational dynamics between users and analytical systems~\cite{setlur2022you}. Zhou et al.~\cite{zhou2024epigraphics} contributed to this area of research by developing an infographics authoring tool based on user-specified messages. 

\hl{Recent advancements in LLMs have ushered in a new era of enhanced support for natural language-based visualization and visual analytics. For instance, Zeng et al.~\cite{zeng2024advancing} optimize multimodal LLMs for chart question-answering, while ChartGPT~\cite{chartgpt} uses LLMs to generate charts from abstract natural language descriptions. Wang et al.~\cite{wang2024aligned, wang2024dracogpt} further explore whether LLMs exhibit perceptual awareness akin to humans. Moreover, various prototypes, both general-purpose~\cite{zhao2024lightva, zhao2024leva} and domain-specific~\cite{yan2024knownet, kim2024phenoflow, liu2024smartboard}, leverage the extensive knowledge base and high configurability of LLMs and highlight their potential in visual analytics.}

The role of language in visual analytics also extends, in the other direction, to the generation of narrative visualizations, where textual descriptions and visual elements jointly convey insights more effectively to users. Systems such as DataParticles~\cite{dataparticles} and Calliope~\cite{shi2020calliope} facilitate generating narrative-driven visual stories from data. Other work has explored theories and techniques for using both textual and visual elements in conveying insights~\cite{stokes2022striking,stokes2024s, latif2021kori, badam2018elastic}.

Recent research has also highlighted the nuanced ways in which language can influence analytical takeaways, wherein captions and textual annotations can guide the interpretation of line charts~\cite{kim2021captions}. This is complemented by Setlur~\etal~\cite{setlur2019inferencing}, who investigated inference in underspecified natural language utterances in visual analytics. In addition, tools such as VizFlow~\cite{text-chart-link}, Olio~\cite{olio}, and SlopeSeeker~\cite{slopeseeker} facilitate semantic search within data repositories, while
other research has contributed to understanding the role of language in collaborative data science~\cite{pang2022data,data-science-team-communicate}.

In setting the stage for the work described here, Setlur and Birnbaum~\cite{setlur2024can} underscored the importance of nuanced language in conveying actionable insights from visual analytics and driving effective decision-making. Their work outlined in a preliminary form a design space integrating \semanticcolor{semantics}, \rhetoriccolor{rhetorics}, and \pragmaticcolor{pragmatics} to frame how language can better communicate actionable insights. Our current work considerably elaborates this design space and implements strategies that leverage LLMs for actionable EDA and storytelling.

\subsection{Human-AI Collaboration for Data Analysis}
Human-AI collaboration for data analysis seeks to harness human intuition and machine execution. Gu et al.~\cite{gu2024analysts-understand-ai-ds, gu2024analysts-respond-ai} sought to understand how data analysts interact with AI systems and demonstrated the importance of the transparency, explainability, and reliability of AI systems. Researchers have also proposed guidelines for designing AI applications in data science, stressing how the workflows, interfaces, and interactions of AI tools should adapt to user needs~\cite{wang2019human-ai-ds, mcnutt2023design, Drosos2024ItsLA}. In particular, Li~\etal~\cite{Li2023WhereAW} reviewed the literature and addressed the implications of where and how AI can be integrated into data science workflows. 

Other research has developed advanced techniques for automating various stages of data science workflows~\cite{wang2021autods, cheng-etal-2023-gpt, Yu2024PyGWalkerOA}. AI systems are now increasingly used to generate insights autonomously~\cite{ma2023insightpilot, lin2023inksight}. Notably, \hl{
Tian et al.~\cite{tian2023interactive} propose an interactive text-to-SQL generation pipeline to support data manipulation.} InsightLens builds a pipeline of LLMs to generate and manage insights from data~\cite{weng2024insightlens}. InsightBench~\cite{sahu2024insightbench} further contributes a benchmark for end-to-end business analytics agents.
There have also been efforts to adapt AI for specific domains, such as scientific research~\cite{jun2019tea,jun2022tisane, bar2020automatically} and education~\cite{li2023edassistant}. More recently, HCI researchers have tried to improve steerability and interpretability of LLM assistants in data science~\cite{kazemitabaar2024improving, xie2024waitgpt}, making AI systems more transparent and user-friendly.

Our work extends prior research by introducing a design space for actionable EDA and storytelling. We use this design space as guidance and guardrails, operationalizing it with LLMs to help them accurately interpret and track analytical objects and results, strategically formulate and apply analytical plans, and offer grounded, contextual, actionable insights. This structured approach also improves explainability and reparability, enhancing the human-AI collaboration experience in data analysis.

\subsection{Human-AI Collaboration for Data Storytelling}
Combining visual analytics with narrative techniques often enhances user engagement and comprehension~\cite{segel2010narrative}. To lower the barriers to data storytelling, past research has explored various AI techniques for creating data stories in different forms~\cite{chen2023does, he2024leveraging}. 

DataTales~\cite{sultanum2023datatales}, which leverages LLMs to weave text and charts into a single cohesive story, is a prototypical dashboard-style system combining text and visualization for data storytelling. DataNarrative~\cite{Islam2024DataNarrativeAD} further develops a bi-agent system to emulate the human data storytelling process. \hl{Aletheia~\cite{fu2024data} tackles another aspect of data storytelling by enabling non-technical data storytellers to fact check data claims with the help of LLMs.} Additional work in this area includes AI-driven storytelling frameworks that incorporate user input~\cite{sun2022erato} and feedback~\cite{wu2023socrates} to enhance the relevance and richness of data narratives, or that exploit the structure of the data~\cite{li2024coinsight}. 
 
Another vein of work focuses on automating slide creation. NB2Slides~\cite{zheng2022telling} was the first system to automatically generate presentation slides from Jupyter Notebooks. Subsequent systems improved this approach by enhancing data fact identification~\cite{li2023notable}, incorporating NLP techniques~\cite{wang2023slide4n}, and enabling users to create outlines~\cite{wang2024outlinespark}. As an engaging format, data videos have also garnered increasing attention from researchers interested in automation~\cite{shi2021autoclips,tang2022smartshots,shen2023data,ouyang2024noteplayer,wang2024wonderflow}. Additionally, AI techniques have been applied to unconventional storytelling formats, such as comic-style narratives~\cite{zhao2021chartstory} and scrollytelling~\cite{lu2021automatic}.

The use of AI in data analytics and data storytelling is a rich space of exciting research. Our work focuses on developing a systematic design space for actionable EDA and storytelling and operationalizing this space in \sys\, with the goal of generating higher-quality actionable data analyses and data stories. As a user-centered system, \sys\ also allows users to directly edit data stories or provide guidance for AI refinement. By combining actionable EDA and data storytelling within a single environment, \sys\ provides an approach that is more fully integrated into analysts' workflows.

% \section{A Conceptual Framework for Actionable Data Storytelling}
% \input{sections/03-dimensions-theory}

\section{Formative Interview and Participatory Design Exploration}
% assess the dimensions' validity and informing tool
%To validate and enrich the design space outlined in Section~\ref{preliminary design space}, w
We conducted formative interviews with nine data analysts. We aimed to understand analysts' workflows, considerations, and challenges in actionable EDA and storytelling, as well as to gauge their attitudes toward AI assistance. In addition, we ran a \hl{participatory design exploration} with participants to explore interfaces and interaction techniques in Jupyter Notebook~\cite{kluyver2016jupyter} that would facilitate human-AI collaboration in EDA and storytelling. We focused on Jupyter Notebook because of its popularity among data analysts for EDA. %with recent received much research attention and which we aimed to integrate AI assistance into.

\subsection{Participants} 
We recruited nine participants (\textbf{P1}~-~\textbf{P9}\hl{; four men, four women, and one non-binary individual}) with extensive experience in actionable EDA and storytelling \hl{from X (formerly Twitter) and three Slack workspaces frequented by data analysts}. Participants represented diverse backgrounds: five were professional data analysts from varying industries, such as environmental science, retail, and education; three were PhD students in natural language processing, human-centered design and engineering (HCDE), and visualization; one was a master's student in HCDE. Participants reported a maximum of 10 years and a minimum of 3 years of experience with data analysis (\textit{mean} = 5.2). \hl{We screened participants using an intake survey where they detailed their actionable EDA and storytelling experience, which they later elaborated during interviews. Professional data analysts primarily performed these activities as part of their job, such as drafting environmental policy proposals or creating sales strategies. Graduate students, on the other hand, conducted these types of analyses for research, e.g., quantitatively comparing human-AI interaction designs and advocating for a specific design in research papers.}

\subsection{Study Procedure}
The study comprised two consecutive sessions. First, we conducted a semi-structured interview lasting around 30 minutes, in which we asked about participants' actionable EDA and storytelling workflows, how they derived and communicated actionable insights, and challenges they often encountered. Toward the end of the interview, we explored how analysts envisioned incorporating AI assistance into EDA and storytelling workflows, including the roles they would like the AI to play and the stages at which they would benefit from AI support. \hl{We provide the question list in Supplemental Materials.}

Next, we engaged participants in a 15-minute \hl{participatory design exploration} around interaction techniques and interfaces for human-AI collaboration in Jupyter Notebook. Inspired by McNutt~\etal's design space for AI assistants in computational notebooks~\cite{mcnutt2023design}, we initially presented participants with three interfaces for inputting natural language commands: targeted cell, inline cell, and side panel (see Supplemental Materials). We then sought feedback on the suitability of each interface for invoking different types of AI assistance, such as generating responses in EDA and presenting data stories. Finally, we opened a virtual whiteboard and invited participants to sketch alternative designs and articulate their affordances. \hl{To encourage deeper discussion and inspire design iterations, we, as interviewers, adopted the perspective of a potential user of \textsc{Jupybara} as appropriate, posing questions and scenarios that might arise in their workflows.}

All sessions took place online. Participants were compensated with a \$20 Amazon gift card. We recorded audio and video for all study sessions. \hl{A thematic analysis of the interviews was performed by one of the authors and subsequently refined through discussions with another author. This analysis initially focused on identifying recurring characteristics and challenges in actionable EDA and storytelling workflows. From these findings, we distilled common considerations into a design space, building upon and expanding the framework proposed by Setlur and Birnbaum~\cite{setlur2024can}.} 

\subsection{Findings}
We now summarize the general workflows for actionable EDA and storytelling and some common challenges in the process.

\subsubsection{Workflows for Actionable EDA and Storytelling}
Actionable EDA and storytelling are fluid, integrated, and iterative processes. Its actual workflows vary considerably based on the nature of the dataset, the analyst's domain expertise, initial objectives, and argumentative needs.

\begin{itemize}[leftmargin=0.55cm]
    \item \textbf{Fluidity.} Most participants described their EDA workflows as involving steps such as data cleaning, visualization, transformation, modeling, and hypothesis testing. In data storytelling, they need to organize their findings, verbalize results to highlight actionable insights, and go through revisions. However, these steps are not fixed or linear. As \textbf{P2} noted, ``\textit{There is not a `fixed workflow' for me. The steps I take to analyze a dataset depend on what it has to offer and what my objectives are.}'' Other participants also pointed to factors like ``\textit{surprising findings}'' and ``\textit{guiding analytical questions}'' as influences on the steps they take and the sequence in which they are taken.
    
    \item \textbf{Integration.} In practice, actionable EDA and storytelling are often integrated processes, echoing Gratzl~\etal~\cite{gratzl2016visual} Especially towards the later stages of projects, analysts often cycle between EDA and storytelling. For example, \textbf{P5} typically ``\textit{switch[es] back and forth between coding and slide-making,}'' citing needs to ``\textit{organize findings in [his] presentation}'' and ``\textit{look for more evidence in the data to support claims in the report.}'' By cycling between EDA and storytelling, analysts can methodically document their findings, continuously refine actionable insights, and effectively plan future analyses.
    
    \item \textbf{Iteration.} Actionable EDA and storytelling workflows are typically messy and iterative, echoing Rule~\etal~\cite{rule2018exploration} When exploring a dataset, analysts constantly reformulate their hypotheses, mental models, and actionable insights according to the results they observe. \textbf{P8} shrewdly characterized this process as ``\textit{incrementally learn[ing] about the dataset.}'' During EDA, analysts navigate many data facts and cull out interesting ones to further examine. Sometimes, they reach dead-ends and need to revise their approaches; sometimes, analysis paths cross, leading analysts to revisit previous insights and uncover deeper ones.
\end{itemize}

\subsubsection{Challenges in the Workflow} 
\label{challenges}
As intricate and even delicate processes, actionable EDA and storytelling are not without their difficulties. We now dive into four recurring challenges participants brought up.
% These aspects of fluidity, intertwinement, and iteration indicate that actionable EDA and storytelling are intricate and even delicate processes, posing considerable challenges to human analysts as well as to both interaction design and the design of possible AI assistants. We now dive into four recurring challenges participants brought up.

\begin{itemize}[leftmargin=0.55cm]
\item \textbf{C1: Identifying appropriate analytical strategies.} To answer analytical questions, data analysts engage in a series of operations on the data, such as imputation, filtering, and correlation analysis. We refer to a series of concerted analytical operations as an \textit{analytical strategy}. All participants concurred that leveraging appropriate analytical strategies is crucial for extracting valid and compelling insights. Yet, identifying what analytical strategies to use is often challenging, as the process requires statistical expertise, domain knowledge, and familiarity with the dataset. Further, analytical strategies often need to be tailored to specific questions. For instance, \textbf{P8} admitted to often needing to ``\textit{Google around to decide which tests to run.}'' \textbf{P9} highlighted that the seemingly trivial task of handling missing values depends on the nature of the dataset and standard practices in the field. Another curious aspect of how industry know-how affects analytical strategies lies in the choice of adjustments and normalizations. \textbf{P2}, a basketball fan, shared that his analyses of player performance often involved adjusting for playing time and overall team performance. Together, these examples emphasize the difficulty of coordinating multiple dimensions in determining reasonable analytical strategies.

\item \textbf{C2: Tracking insights and analysis history.} Similar to  Weng \etal~\cite{weng2024insightlens}, we found that tracking insights and analysis history places heavy cognitive burdens on analysts. Every participant agreed that managing insights in EDA is a necessary yet demanding component that pervades the whole EDA and storytelling workflow. \textbf{P1} observed, ``\textit{I pick up findings here and there. It can be challenging to keep track of all of them, considering how many findings I encounter.}'' To document insights, analysts often take notes and screenshots, which provide fodder for insight association and data storytelling. Further, analysts expressed a need to record analysis history, including both the paths that lead to insights and those that result in dead-ends. As \textbf{P9} noted, the fluid and iterative nature of EDA means that the process can be ``\textit{a combination of breadth-first search and depth-first search.}'' Documenting the analytic approaches (i.e., the analytical operations on the data) is not only helpful for informing future analysis and course correction, but also for creating a coherent and persuasive data story. Nonetheless, due to the potentially large number of steps taken to analyze the data, documenting them becomes so time-consuming and mentally taxing that most participants do not engage in this practice systematically, instead relying on memory to recall their analysis paths.

\item \textbf{C3: Finding the right language and narrative structure to effectively convey actionable insights.} 
All participants reported experience in presenting actionable insights in slide decks and written reports. Analysts widely agreed that the language and narrative structure used to verbalize findings significantly impact how the audience perceives them. This is especially true for actionable insights, which inherently carry persuasive intents. Yet, drafting effective actionable data narratives is often a challenging exercise. At the ``lowest'' level, analysts must deliberate word choices when conveying their results. \textbf{P4} observed that ``\textit{there are a million ways you can describe your results, and each has its own meaning.}'' Choosing the right language, then, becomes a matter of ``experience'' and ``gut feeling.'' At a ``higher'' level, analysts need to determine which results to highlight and the appropriate level of detail to provide. These decisions, in turn, depend on a range of factors, such as the prospective actionable insights, the background of the audience, and the context in which the data story is presented, complicating the process of crafting an effective narrative.

\item \textbf{C4: Leveraging relevant domain knowledge to derive actionable insights from data facts.}
Actionable insights do not exist in a vacuum. In order to transform raw data facts from EDA into actionable insights, analysts need to contextualize the results and justify their proposed courses of action by identifying and applying relevant domain knowledge. It can be overwhelming to sift through the vast body of external knowledge required to find the most relevant information. This challenge is particularly pronounced when analysts work across multiple domains or with unfamiliar datasets. According to \textbf{P7}, researching pertinent domain knowledge to augment the narrative is ``\textit{usually harder than getting the results from data analysis.}'' For domains he was unfamiliar with, he often struggled to find relevant sources to augment the narrative. Moreover, even when relevant domain knowledge is identified, analysts need to carefully reason through how to apply it. The particularities in each dataset require analysts to meticulously evaluate how domain knowledge intersects with their data findings. As \textbf{P6} aptly put it, ``\textit{getting results [in EDA] is not the end---it's just the beginning.}''

\end{itemize}
.

\subsection{A Three-Dimensional Design Space}
To map the design space of actionable EDA and storytelling, we asked participants to describe strategies they used to mine and communicate actionable insights and to provide examples. 
\hl{The first author, in discussion with the other authors, then coded these strategies.} 
We now present this design space in further detail and illustrate how each dimension manifests in both EDA and storytelling. We will also show how our design space subsumes the challenges identified in Section~\ref{challenges} \hl{and discuss them in the context of relevant literature}.

\subsubsection{\semanticcolor{Semantic Dimension}}
\label{semantic}
The \textcolor{DeepTeal}{semantic dimension} involves precisely specifying the analytical objects (i.e., \textit{what} is being analyzed) and interpreting and tracking the results. It is through language that analysts translate these objects and results into semantic properties they can reason with and about, and convey their nuanced implications with precision to readers. As such, \semanticcolor{semantic precision} precedes and underpins the generation of insights. \hl{This formulation aligns with existing literature, which also emphasizes the importance of how meaning is assigned to data~\cite{grolemund2014cognitive} and how findings from EDA are articulated in a manner that preserves the integrity of the analysis \cite{qiao:2021}.}

\textbf{EDA.} A thorough understanding of the data attributes is essential for any analysis on a dataset. All participants alluded to grasping the semantics of the analytical objects as one of the initial steps in EDA. For instance, analysts often ask themselves: what attributes exist in the dataset? What are their data types? Answering these questions helps analysts develop a clearer sense of \textit{what} is being analyzed and the gamut of analytical questions the dataset can possibly support. To enhance the semantics of a dataset, analysts can further add metadata such as descriptions and data provenance, or join the data with other datasets. \hl{Kandel et al.'s seminal interview study highlights that the seemingly straightforward task of understanding data is, in reality, riddled with challenges~\cite{kandel2012enterprise}. This has led to the development of various tools aimed at supporting tasks like data profiling~\cite{epperson2023dead}, table profiling~\cite{huang2024cocoon}, and data difference identification~\cite{sutton2018data}. Despite their differences, these tasks share a common goal: providing an accurate semantic grounding of the data to establish a solid foundation for subsequent analysis.}

In addition to defining the semantics of analytical objects, the \textcolor{DeepTeal}{semantic dimension} encompasses ensuring the analytical results carry valid semantics. For instance, in order to glean insights from a visualization, analysts first need to make sure that the visualization is an honest representation of the data, since inappropriate design choices can distort the true patterns in the data and lead to unsound insights downstream. \textbf{P7}'s remark best captures a common approach to this: ``\textit{Every time the code gives me output, I immediately check if the results make sense, and I often catch silly mistakes like using the wrong features in my model.}'' \hl{Prior empirical studies have similarly observed the critical role of verifying the semantic validity of results for data scientists, both in traditional human workflows and human-AI collaboration systems~\cite{gu2024analysts-respond-ai, gu2024analysts}.} Another critical aspect of the \semanticcolor{semantic dimension} is keeping track of analytical results (echoing \textbf{C2}). Figuratively, analysts construct a ``semantic web'' of results to track and associate data facts, which they grow as they uncover new findings. This ``semantic web'' not only informs subsequent steps in EDA, but also provides raw material for insights, forming a common substrate for both EDA and storytelling. \hl{And it is exactly for this reason that many tools emphasize helping analysts manage and track their findings (e.g., ~\cite{weng2024insightlens}, ~\cite{wang2022interactive}, ~\cite{eckelt2024loops}).}

\textbf{Data Storytelling.} While an accurate \textit{conceptual} understanding of the semantics of analytical objects and results is generally sufficient in EDA, writing a data story further requires analysts to find \textit{proper wordage} to express these semantics, echoing \textbf{C3}. In this sense, data storytelling entails articulating the semantics that are constructed and curated in EDA. \hl{Consistent with our findings, prior research suggests that the first step in data storytelling is ensuring that language accurately represents the trends, anomalies, and relationships present in the data~\cite{segel2010narrative,riche2018data}.} 
When determining the strength of a correlation, for instance, analysts understand that an \textit{r} value of 0.7 indicates a relationship, but they must decide whether to describe the value as ``moderate'' or ``moderately strong.'' (In some circumstances, this issue can be quite fraught: analysts in the US intelligence community, for example, have developed strict guidelines on appropriate numeric ranges for terms such as ``likely.'') \textbf{P6} gave another example on presenting parameter estimates: while a 95\% confidence interval in frequentist terms suggests the range would capture the true parameter in 95\% of repeated studies, a 95\% credible interval in Bayesian analysis indicates a 95\% probability that the true parameter lies within that range. 

In many cases there are few established guidelines on how to characterize results; analysts must exercise even more caution in choosing the right language. For instance, upon seeing a sharp decline in sales on a line chart, analysts conceptually understand the drop but need to choose the appropriate wording, such as ``fall,'' ``decline,'' or ``crash,'' to precisely convey the extent of the change. \hl{This careful selection of language echoes recent work by Bromley and Setlur~\cite{bromley2023difference}, who demonstrate that the semantics behind language influence human perception and understanding of data patterns via an example of how visual features in line charts are associated with different natural language trend descriptors.} Another common strategy for \semanticcolor{semantic precision} is to use domain-specific language. For example, a flat trend in the financial sector might be described as ``steady,'' whereas in weather forecasting, ``unchanged'' would be a more suitable term. In summary, these examples demonstrate the power of language in precisely communicating the nuances of analytical results---and the need to be careful in doing so.

\subsubsection{\rhetoriccolor{Rhetorical Dimension}}
The \textcolor{RoyalBlue}{rhetorical dimension} involves deploying analytical strategies to derive compelling results and orchestrating the narrative with an eye toward generating, bolstering, and advancing actionable insights. \hl{In a sense, this dimension resembles the use of \textit{rhetoric} to support actions or responses commonly studied in argumentation~\cite{perelman:1969}:} it ensures the trajectory of EDA and the presentation of analytical strategies and results are effectively geared toward persuasively conveying the insights. Hence, the \rhetoriccolor{rhetorical dimension} subsumes the \semanticcolor{semantic dimension} and supports the \pragmaticcolor{pragmatic dimension}.

\textbf{EDA.} Much like rhetorical devices in persuasive writing, analytical strategies in EDA serve vital persuasive purposes. If the semantics of analytical results provide evidence for the insights, then it is through carefully chosen strategies that analysts surface the most relevant findings in EDA (echoing \textbf{C1}). \textbf{P1} and \textbf{P4} pointed out that while there could be multiple viable analytical strategies for a given task, there are often nuanced differences in the perspectives they underscore. Consider choosing a dimensionality reduction method: selecting principal component analysis over t-SNE emphasizes the preservation of variance, which could be more effective when arguing for the importance of certain features. When selecting a method for reliable long-term time series forecasting, ARIMA is usually preferred over exponential smoothing for its superior ability to account for trends and seasonality. As another example, choosing between simple correlation and partial correlation can shape how relationships between variables are perceived. Partial correlations, which control for other variables, are particularly useful when arguing for the independent effect of a variable. \hl{Interestingly, such rhetorical considerations align with a recurring theme in multiverse analysis: reasoning about multiple potential paths of analysis~\cite{liu2020boba, sarma2023multiverse}. Depending on the analysis objectives~\cite{schweinsberg2021same} and the application domain~\cite{jung2022domain}, analysts must carefully evaluate analytic plans and execute them faithfully---a task that presents its own set of challenges~\cite{li2023edassistant}.}

In addition to deciding which analytical strategies to employ, analysts must document the strategies with which they have experimented during the analytic process (echoing \textbf{C2}). By tracking analytical results (\semanticcolor{semantic dimension}) and strategies (\rhetoriccolor{rhetorical dimension}), analysts can better understand the analytical paths taken, recognize dead ends, expose unexplored questions, and revisit potential blind spots (\textbf{P8}). \hl{Recognizing this challenge, several papers have proposed tools to track and document analytic decisions~\cite{kery2019towards, wang2022documentation}, though they still demand considerable cognitive effort from analysts.} Insofar as the analytical strategies determine which data facts are revealed and, therefore, which actionable insights are derived, the \rhetoriccolor{rhetorical dimension} critically shapes the direction of EDA. 

\textbf{Data Storytelling.} An effective data story is not merely a compilation of data facts---strategically curated and presented information often resonates more powerfully with the audience (echoing \textbf{C3}). To begin with, analysts must determine which analytical results from EDA to include in the data story, with the aim of identifying a set of data facts that most effectively supports the take-home messages. While it is tempting to include only findings that support the desired narrative, acknowledging contradictory or unexpected results can sometimes enhance the credibility of the story and provide a more balanced perspective. Next, as \textbf{P6} highlighted, analysts need to decide the order in which to present these findings. A logical and coherent presentation of data facts moves readers ineluctably toward the main conclusions. In this regard, selecting the right connectives with which to convey analytical results is essential for weaving the findings together cohesively. Transitional phrases like ``as a result,'' ``in contrast,'' and ``surprisingly'' elucidate logical connections between data findings and keep the audience engaged. Finally, analysts often need to explicitly narrate the analytical strategies used. Over half of the participants expressed that doing so not only clarifies the methodology but also reinforces the validity of the insights\hl{, a perspective supported by Hullman et al.~\cite{hullman:2011}}. It is also important that the level of detail aligns with the technical background of the audience---tech-savvy readers may appreciate detailed explanations, such as why a particular regression model was chosen, while others might prefer a broader overview. \hl{While many papers have pointed this out~\cite{wongsuphasawat1911goals, segel2010narrative, sauer2016audience}, perhaps Drucker et al. summarize it the best---virtually every aspect of how one should communicate data findings ``depends on the audience.''~\cite{drucker2018communicating}}

Besides structural considerations, multiple participants mentioned that thoughtful word choices could also enhance the persuasive power of a data story. For instance, while there may exist multiple accurate word choices to describe the same result, each can carry subtly different overtones: in the context of stock prices, ``crash'' and ``fall sharply'' both describe a rapid decline, but the former implies a more severe, potentially irrevocable impact, endowing the word with greater persuasive power to prompt stakeholders to action. To give another example, while ``stagnant'' and ``stable'' share similar meanings, they evoke entirely different expectations---``stagnant'' typically carries a negative connotation, suggesting a lack of growth, whereas ``stable'' implies consistency and reliability, which is generally viewed more positively. To sum up, through structural cohesion and lexical nuance, data stories can better communicate the desired significance and implications of actionable insights. 

\subsubsection{\pragmaticcolor{Pragmatic Dimension}}
The \textcolor{BurntSienna}{pragmatic dimension} involves augmenting analytical results with relevant external knowledge to suggest effective courses of action\hl{, thereby connecting data insights to real-world decision-making~\cite{Kahneman2013,keeney:1992}}. Building upon precisely conveyed analytical results and carefully chosen analytical strategies, the \pragmaticcolor{pragmatic dimension} culminates in grounded, actionable insights. Whereas the \semanticcolor{semantic} and \rhetoriccolor{rhetorical} dimensions manifest differently in EDA and storytelling, the distinction is much blurrier for the \pragmaticcolor{pragmatic dimension}. Therefore, we address EDA and storytelling together here.

Despite its significance, the concept of insight has been defined in varying ways in the literature. Over time, however, scholars have increasingly moved from viewing insight as mere data facts~\cite{choe2015characterizing} to embracing a more sophisticated perspective that integrates analytical results with domain knowledge~\cite{battle2023we}. This more nuanced view is all the more necessary when discussing actionable insights, since effective decision-making in the wild must contextualize data findings in domain knowledge. In practice, actionable insights can take many forms, such as performance improvement, predictive analysis, and decision support~\cite{schneider2014pragmatics}. \textbf{P3} gave an illustrative example from his consulting services: upon observing stagnation in market growth, particularly among younger demographics, he suggested targeted marketing campaigns on social platforms like TikTok and Instagram. In this case, external knowledge about the influence of popular social media platforms on younger audiences helped him connect data findings to practical solutions. Furthermore, past experiences and domain knowledge must often be adapted to the specific situation. \textbf{P3} recounted another case where his team advised a client to replicate successful marketing campaign strategies from the U.S. in Asian markets, which initially failed to resonate with the Asian audience. Only after accounting for cultural differences and revisiting key assumptions did they see a significant improvement in conversion rates. To give yet another example, while insights presented to senior executives might focus on high-level strategic implications, insights shared with operations teams more likely emphasize implementation details. Although challenging on many fronts, the organic combination of data facts with domain knowledge is essential for generating practical and actionable insights (echoing \textbf{C4}).

\hl{Notably, there is a significant gap in research addressing the} \pragmaticcolor{pragmatic dimension} \hl{of EDA and storytelling. Although a key goal of visualization systems is to assist decision making, a review of visualization papers at leading venues highlights that decision-related tasks are seldom represented in current visualization task taxonomies, requiring the field to prioritize supporting practical decision-making processes~\cite{dimara2021critical}. In the broader scope of data science, Crisan et al.'s~\cite{crisan2020passing} literature review reveals a lack of focus on situating analysis results in terms of real-world applications as well. For example, while the prominent standard for industry data mining, CRISP-DM~\cite{wirth2000crisp}, consists of a ``deployment stage'' in which data practitioners generate an actionable analysis report or implement a repeatable data mining process, most studies applying this framework do not include this phase~\cite{schroer2021systematic}. We hope our formulation of the} \pragmaticcolor{pragmatic dimension} \hl{inspires further research into effectively integrating EDA and storytelling with real-world applications.}

\subsection{Design Considerations for an AI Assistant for Actionable EDA and Storytelling}
Most participants expressed enthusiasm for an AI assistant for actionable EDA and storytelling, citing its ability to automate coding and leverage vast domain knowledge for deeper insights. However, they also raised concerns, such as the risk of hallucination and challenges in verifying the analysis. In order to empower actionable EDA and storytelling with the generative power of LLMs while mitigating associated risks, we distill five design goals from our formative interviews:

\textbf{D1. Integration into analysts' existing EDA and storytelling workflows.} We aim to build a system that respects analysts' existing actionable EDA and storytelling workflows. For easier uptake, our system should build on tools and environments familiar to analysts. Given the tight coupling of EDA and storytelling in communicating actionable insights, we also want our system to support both functions. In addition, we require our system to allow smooth cross-referencing between EDA scripts and data stories.

\textbf{D2. Optimization for the design space.} Our design space for actionable EDA and storytelling outlines key dimensions an AI assistant can optimize for. Moreover, these dimensions subsume the challenges in EDA and storytelling: optimizing for the \semanticcolor{semantic} and \rhetoriccolor{rhetorical} dimensions jointly tackles \textbf{C1} through \textbf{C3}, and doing so for the \pragmaticcolor{pragmatic dimension} addresses \textbf{C4}. Nonetheless, to effectively bring this theoretical framework into practice, our tool needs to adopt effective strategies to operationalize the design space.

\textbf{D3. Steerability.} Our interviews revealed analysts' need for control over the AI assistant’s behavior. For instance, analysts preferred using the AI as an executor when they had clear analysis plans but desired more proactive, agentic involvement from the AI when they were uncertain or lacked direction. Therefore, we aim to provide a steerable system, one that can accurately interpret user intent and undertake the appropriate level of agency. 

\textbf{D4. Explainability.} While participants were impressed by the generative capabilities of LLMs, most expressed reservations about their lack of transparency. To address this, we will ensure our system offers explanations to enhance user understanding and trust in both EDA and storytelling. Additionally, we aim to provide a feature that allows users to engage in threaded conversations with the AI for clarification on its decisions and interpretations.

\textbf{D5. Reparability.} In many cases, analysts may want to repair the AI's responses. We will strive for two forms of reparability: direct manipulation and user-guided AI refinement. In direct manipulation, analysts manually adjust the AI's output. In user-guided AI refinement, analysts provide instructions, and the AI implements the changes accordingly. These mechanisms ensure flexibility and control, enabling analysts to refine AI contributions as needed.

\section{\sys}

We now introduce \sys, an LLM-based multi-agent AI assistant for actionable EDA and storytelling. \sys\ is a Jupyter Notebook extension, embedding the AI assistant within an environment familiar to data analysts~(\textbf{D1}). Compared with AI assistants accessible through custom interfaces such as ChatGPT and \textit{InsightLens}~\cite{weng2024insightlens}, all but one participant in our formative interviews preferred accessing such a system through Jupyter Notebook, as the platform retains their ability to author code directly, allows for easy editing of AI-generated content, and obviates the need for switching interfaces when coding. Recent work (\eg~\cite{li2023notable}, \cite{zheng2022telling}, \cite{wang2024outlinespark}) has also highlighted the benefits of computational notebooks in supporting data storytelling, particularly through offering a contextualized storytelling experience with easy access to data findings.

\sys\ offers a natural language interface for automatic EDA and storytelling and adopts an agentic workflow. Utilizing design-space-aware prompting and multi-agent architectures, \sys\ effectively operationalizes our proposed design space~(\textbf{D2}). \sys\ also allows for easy steering~(\textbf{D3}): analysts can express their analytic intent with varying degrees of specificity and complexity, and the system will strive to accurately interpret the intent and respond with the appropriate level of agency. By presenting analysis plans, code comments, and interpretations in EDA, providing a dedicated tab for clarification, and displaying tooltips for explanations in storytelling, \sys\ aims for transparency and explainability~(\textbf{D4}). In addition, \sys\ supports both direct manipulation and user-guided AI refinement of AI-generated content, providing reparability~(\textbf{D5}).

\subsection{Interfaces and Functionalities}
The interface of \sys\ comprises two panels (Figure~\ref{fig:teaser}). The left panel features a canonical Jupyter Notebook augmented with an AI copilot for EDA. The hideable right panel (hereafter referred to as the ``side panel'') allows users to tune system settings, engage in threaded conversations with the AI for clarification, track insights, and generate and refine data stories with AI support. The interface adopts a tabbed design to separate these features and avoid clutter. The two-panel layout of the interface allows users to cross-reference both panels with the Notebook as an anchor.

\subsubsection{EDA Copilot}
In the Notebook, users can invoke the help of \sys\ via natural language. To do so, they create a new cell, input their instructions, and activate the AI through a button in the cell toolbar (Figure~\ref{fig:load dataset}(A)). The LLM then responds agentically in the cell(s) below (Section~\ref{Agentic Behavior in EDA}). Users can interrupt the AI execution at any time by pressing a stop button. Two modes are available for EDA: single-agent and multi-agent. In the single-agent mode, one LLM alone addresses the user query; we discuss the multi-agent mode in detail in Section~\ref{multi}. \hl{While the single-agent mode is faster, more cost-effective, and suitable for simpler tasks, the multi-agent mode is better equipped for complex queries. }

\subsubsection{Settings Tab}
The \textit{Settings} tab allows users to configure the settings of \sys\ (Figure~\ref{fig:load dataset}(B)). Currently, \sys\ enables users to choose between single- and multi-agent modes for both EDA and storytelling. For each agent, users can select between GPT-4o and Claude~3.5~Sonnet.

\subsubsection{Clarification Tab}
During EDA, analysts may have various questions about AI-generated responses. While they could create a new cell to query the system, this approach may disrupt the flow of the Notebook, as some clarifying content (\eg questions about Python syntax) might not directly contribute to the analysis. \hl{Prior work has demonstrated that separating side queries and clarification questions from main queries into a dedicated ``side bar'' can be an effective approach~\cite{kazemitabaar2024improving}.} Hence, we adopt a design where we treat each cell as a thread and allow users to select any cell in the Notebook to engage in a threaded conversation with the AI in a tab on the side panel (Figure~\ref{fig:clarify}(B)). When a user selects a cell and issues a query related to that cell, the user query, the selected cell, and the entire Notebook are passed to an LLM to address the question. This approach provides the requisite context to the LLM, while more cleanly separating analytical questions and clarifying questions. 

\subsubsection{Insights Tracking Tab}
As computational notebooks increase in length, analysts find it progressively more challenging to keep track of their insights~\cite{weng2024insightlens}. To address this challenge, we incorporate a side panel tab that leverages an LLM to automatically summarize key insights (Figure~\ref{fig:insights}). Recent research on insights suggests that the most valuable insights are not merely data facts, but also include the provenance of these facts and the domain knowledge used to contextualize and augment them~\cite{battle2023we}. To operationalize this definition, we prompt an LLM in a Chain-of-Thought manner~\cite{wei2022chain}, guiding the model to first organize the Notebook by analytical questions and then outline the analytical objects, operations, data facts, and domain knowledge involved in each question. \hl{Inspired by prior work on visualizing multiverse analysis, model architecture, and LLM responses~\cite{liu2020boba, jiang2023graphologue, suh2023sensecape, wongsuphasawat2017visualizing}, we opt to render insights as directed acyclic graphs (DAGs) due to their intuitive affordance. The DAG formulation also naturally integrates various elements in Battle et al.'s framework for insights~\cite{battle2023we}. Specifically, we instruct the LLM to represent the aforementioned ingredients using the \texttt{Mermaid} library as a DAG,} where the nodes represent analytical objects, data findings, or external knowledge, and the edges represent analytical operations. Nodes are also color-coded: yellow nodes are analytical objects or data findings derivable from the dataset, and green nodes correspond to external knowledge that informs the analysis. The graph also serves as an interactive index for the Notebook---users can click on nodes or edges, triggering an LLM query that identifies and scrolls to the most relevant Notebook cell, streamlining navigation and recall of the analytical process. 

\subsubsection{Data Storytelling Tab}
The \textit{Storytelling} Tab on the side panel is the hub for data storytelling (Figure~\ref{fig:generate data story}). Here, users can provide information about how to generate the data story (\eg who the target audience is); \sys\ will then produce a data story (in either single- or multi-agent mode) as an HTML page based on the analyses in the Notebook, which users can simply deploy online or export as a PDF document. The data story highlights sections in three different colors, each corresponding to one of the three dimensions. When users hover over the highlighted text, tooltips appear, explaining the language choices or the bases for the insights. As a user-centered system, \sys\ allows users to easily edit AI-generated data stories, either manually in a live, side-by-side HTML editor (Figure~\ref{fig:editor}), or by offering feedback and delegating revision to the AI (Figure~\ref{fig:ai edit}). Users can provide ``Global Feedback,'' which applies to the entire data story (such as adjustments to the writing style), or ``Local Feedback,'' which targets specific user-selected text. Based on the feedback, the original data story, and the Notebook content, an LLM revises the story accordingly.

\subsection{Agentic Behavior in EDA}
\label{Agentic Behavior in EDA}
An \hl{agentic} EDA assistant in a computational notebook \hl{should} be capable of generating diverse content, including analytic plans, code, and interpretations, at the appropriate times. Moreover, responses to complex analytical queries might entail generating multiple types of content sequentially. To achieve this functionality, we prompt the backing LLM in \sys\ following the ReACT paradigm~\cite{yao2022react}, instructing the model to decompose complex queries into steps, respond with outputs at each step, and observe their effects. We further treat each Notebook cell as a unit of response. Each time the LLM produces a response, the system must specify whether the response should be placed in a code cell or markdown cell before it is appended to the Notebook. The system then executes the cell and sends the results (if any) back to the LLM, which then decides whether further actions are needed. This process is repeated until the LLM deems the original query to be sufficiently addressed.

This agentic workflow is also well-suited for simple queries: when the LLM recognizes that the user’s query has been sufficiently addressed by the initial response, the system can opt not to follow up, handing control back to the user. Similarly, this approach handles queries with varying levels of specificity effectively. Given detailed instructions, \sys\ functions as an executor grounded in the plan provided by the user, whereas for less specific queries, the ReACT paradigm helps yield nuanced responses via multi-step reasoning. This flexibility of \sys\ provides a significant degree of user steerability (\textbf{D1}).

\begin{figure*}[t]
  \centering
  \includegraphics[width=\textwidth]{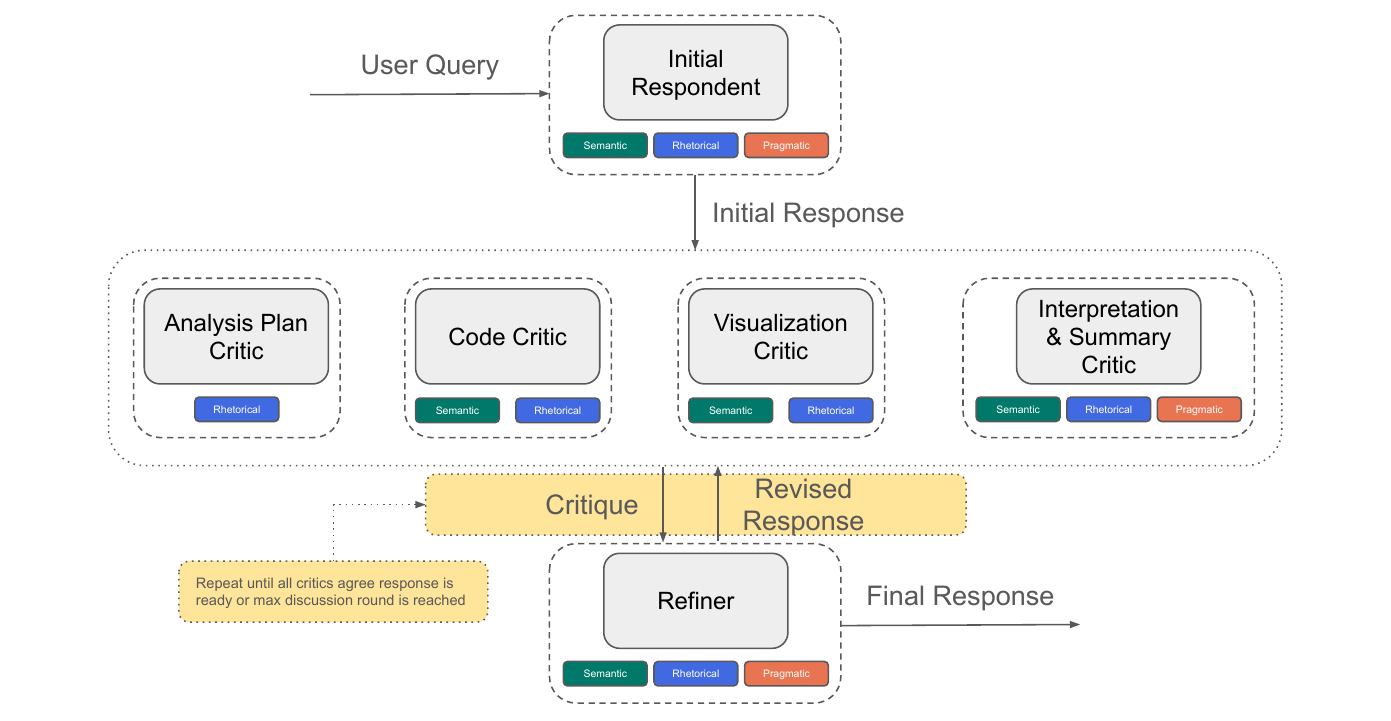}
  \caption{The multi-agent architecture for EDA in \sys. Given a user query, an \textit{Initial Respondent} provides an initial response, which is then critiqued by four \textit{Critics}. Based on the critiques, the \textit{Refiner} improves the response and sends the revised version back to the \textit{Critics} for review. The discussion between the \textit{Critics} and the \textit{Refiner} continues until all \textit{Critics} agree the response is ready or a defined maximum number of discussion rounds is reached, at which point the final response is presented to the user. Notably, each dimension of the design space is addressed by at least three agents, potentially enhancing the quality of the response.}
  \label{fig:eda arch}
\end{figure*}

\subsection{Operationalizing the Design Space with LLMs}
While the design space outlines important dimensions for actionable EDA and storytelling, we still need to operationalize the space---i.e., we must translate theoretical considerations in the framework into concrete guidelines concerning how to develop actionable insights. It is important to note that there can be many strategies for operationalizing the design space. Here we discuss two strategies we utilize in \sys: design-space-aware prompting and multi-agent architectures. 

\begin{figure*}[t]
  \centering
  \includegraphics[width=\textwidth]{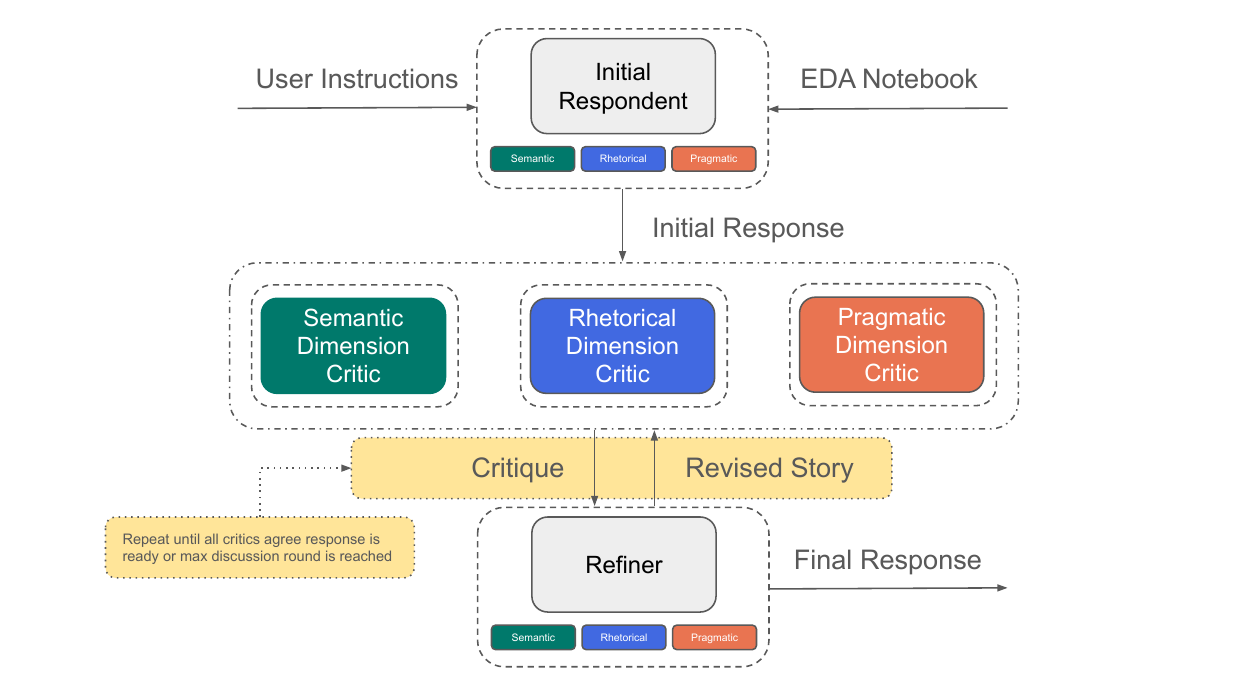}
  \caption{The multi-agent architecture for data storytelling in \sys. An \textit{Initial Respondent} generates a first draft of the data story, which is then critiqued by three \textit{Critics}, each focusing on one dimension of the design space. The \textit{Refiner} then improves the response and sends the revised version back to the \textit{Critics} for further review. This process continues until all \textit{Critics} agree the response is ready or the maximum number of discussion rounds specified is reached, at which point the final response is sent to the user. Notably, each dimension of the design space is addressed by three agents, potentially enhancing the quality of the response.}
  \label{fig:story arch}
\end{figure*}

\subsubsection{Design-Space-Aware Prompting}
General-purpose LLMs, such as GPT-4 and Claude~3.5, are pretrained on open-domain corpora and instruction-tuned for following directions. While they are capable of handling user queries in EDA and generating data stories, their responses can reflect patterns in the training data that are flawed or contextually inappropriate. To provide guidance and guardrails to LLM, the system prompts we utilize are formulated with considerations from our design space.

In EDA, the LLM needs to generate both natural language (\eg analysis plans, interpretations) and code (\eg visualizations, data cleaning scripts). These varied types of content, however, do not map one-to-one to the three dimensions. For instance, a markdown cell could contain interpretations of results (\semanticcolor{semantic dimension}), analysis plans (\rhetoriccolor{rhetorical dimension}), or actionable insights (\pragmaticcolor{pragmatic dimension}). Directly providing the LLM with the definitions of the three dimensions seems too abstract and broad-brush to enable meaningful engagement with the design space in such a flexible setting. Instead, we distill a set of concrete guidelines from each dimension that the LLM should follow. For example, for the \semanticcolor{semantic dimension}, we instruct the LLM to ``always interpret statistical results and visualizations'' if LLM-generated cells contain them. For the \rhetoriccolor{rhetorical dimension}, we prompt the model to ``keep the user in-the-loop by telling them your plans.'' To inform better choices of analytical strategies, we further encourage the LLM to generate visualizations before conducting statistical tests to understand the semantics of the data. Thus, although not directly instructed with the definitions of the dimensions, the LLM adheres to practices that materialize these design considerations.

In data storytelling, \sys\ focuses on generating only one type of content: a (largely) natural language narrative communicating data findings and actionable insights. This relative homogeneity makes data storytelling with LLMs more amenable to direct operationalization of the design space. We provide definitions for each dimension of the design space, along with examples of how they manifest in data stories. For the \semanticcolor{semantic dimension}, for example, we explicitly instruct the LLM to deliberate how to accurately ``convey important results of the analysis'' and include examples such as the one illustrating contextually relevant trend descriptors in Section~\ref{semantic}. We include all the prompts in Supplemental Materials.

\subsubsection{Multi-Agent Architectures}
\label{multi}
Even with the design space as guidance or guardrails, LLMs might still overlook important details in their initial responses, not least because of the challenges in accounting for the extensive set of guidelines derived from our design space. 
\hl{Recent studies have leveraged multi-agent architectures to improve LLM reasoning (\eg~\cite{Du2023ImprovingFA}, \cite{Zeng2022SocraticMC}, \cite{Li2022ComposingEO}), where agents interact to iteratively enhance responses. The potential of this approach has not yet been fully realized in the realm of EDA and storytelling. While InsightLens~\cite{weng2024insightlens} employs multiple agents for generating and tracking insights, its approach is better described as a multi-agent \textit{pipeline}, where Agent 1 responds to the user query, Agent 2 extracts insights from Agent 1's output, and Agent 3 organizes Agent 2's insights. This one-way pipeline limits inter-agent interaction, which could be crucial for addressing complex queries requiring iterative refinement; if any agent makes a mistake, the other agents cannot step in and help correct it. To overcome these limitations, we propose two multi-agent architectures, tailored separately for EDA and storytelling. Unlike the single-agent mode, in which each user query is handled by a single LLM, the multi-agent mode involves multiple agents collaborating to deliver more nuanced and reliable results. To better scaffold multi-agent collaboration, we further construct the multi-agent architecture with the design space in mind.~\footnote{\hl{We note that concurrent work, DataNarrative~\cite{Islam2024DataNarrativeAD} (first uploaded to arXiv in August 2024), is the closest technique to our approach. Their method adopts two agents to generate short data narratives for charts, where one reflects and one verifies. Our multi-agent architectures not only cover EDA on top of data storytelling, but also harness a design space to guide generation, critique, and refinement.}}}

Responses to EDA queries can be broadly divided into three categories: analysis plans, code, and interpretations \& summaries. Since it may be difficult for a single LLM to effectively factor in all the guidelines from the design space, we introduce specialized \textit{Critics} to review the responses, evaluate whether the current response is ready, and generate critiques (if any). In addition to assigning agents for analysis plans, code, and interpretations, we designate another agent specifically for visualizations. Although visualizations are technically generated with code, their rich design considerations warrant a separate \textit{Critic}. The advantage of this architecture is that each \textit{Critic} only needs to reason over a much smaller set of considerations, potentially enabling better identification of gaps or oversights in the initial response. Additionally, we introduce the \textit{Refiner}, an agent tasked with refining the initial response based on the critiques provided by the \textit{Critics}. 

\begin{figure*}[t]
  \centering
  \includegraphics[width=\textwidth]{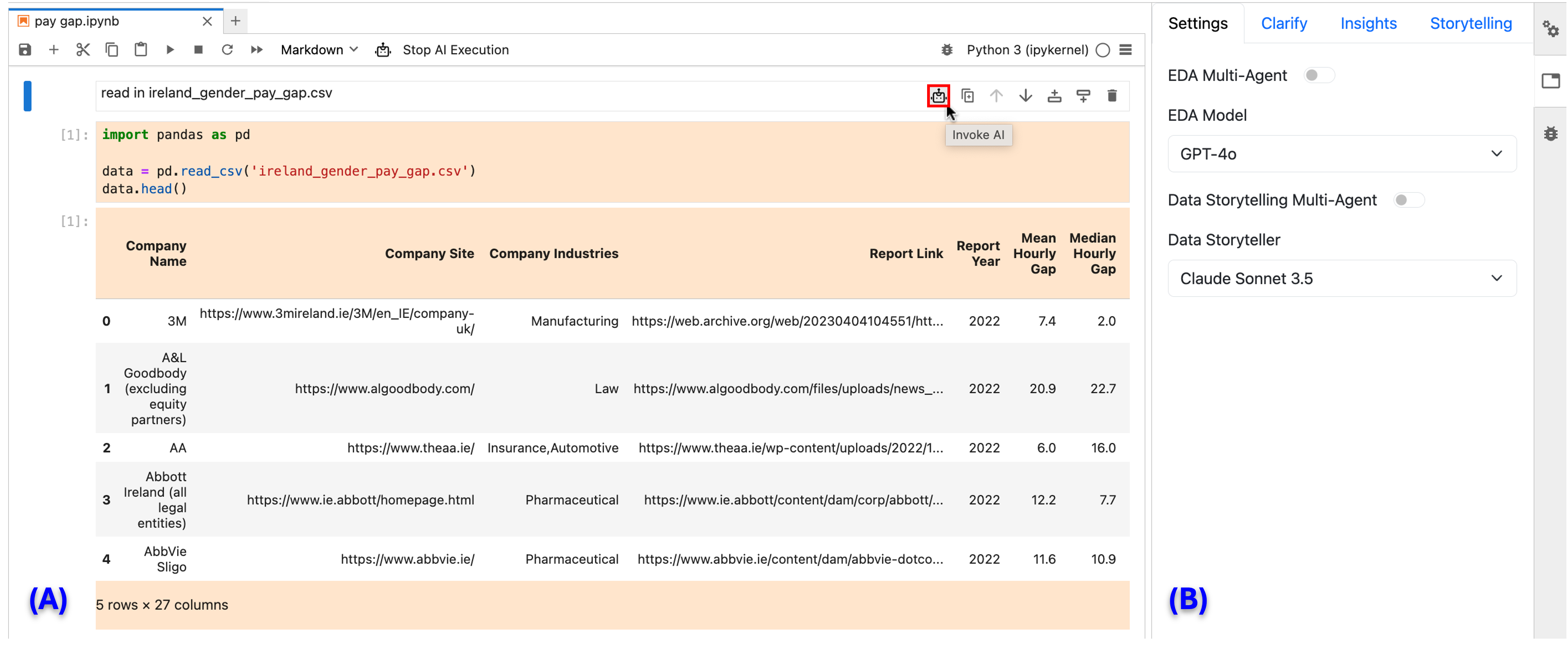}
  \caption{(A) Rae inputs her command in a Notebook cell and clicks the ``Invoke AI'' icon in the cell toolbar. \sys\ inserts its response in a new Notebook cell and executes it, fulfilling the request. (B) Rae can configure the settings of \sys\ in the \textit{Settings} tab. For each LLM agent in the system, Rae can choose between GPT-4o and Claude~3.5~Sonnet.}
  \label{fig:load dataset}
\end{figure*}

Figure~\ref{fig:eda arch} illustrates the multi-agent workflow for EDA. Given a user query, an \textit{Initial Respondent} (the same agent handling user queries as in the single-agent mode) first generates an initial response. The four \textit{Critics}, each focusing on one of analysis plans, code, visualizations, and interpretations \& summaries, then independently critique this response. Each critic is prompted following the Chain-of-Thought paradigm~\cite{wei2022chain} to first summarize existing content in the Notebook for a better understanding of the context before evaluating the response. Importantly, each \textit{Critic} is instructed to decide whether the next response should pertain to its area of focus based on the user query and content in the Notebook. If so, the agent then evaluates the response based on the provisioned considerations and its knowledge of general best practices. If not, the agent refrains from providing input. This approach ensures that, even if the initial response is code, for example, the \textit{Analysis Plan Critic} can intervene and request that a plan be generated before proceeding with the code. Next, the critiques are aggregated and passed to the \textit{Refiner}, which first decides which critiques to accept and then refines the response accordingly. For each rejected critique, the \textit{Refiner} provides a rationale. The refined response and the rationales are then sent back to the \textit{Critics} for another round of review. This iterative process continues until all \textit{Critics} deem the response acceptable, or a preset limit on discussion rounds is reached, at which point the response is returned to the user.

Our multi-agent system enhances the operationalization of the design space by engaging multiple agents to iteratively refine along each dimension. Both the \textit{Initial Respondent} and the \textit{Refiner} need to implicitly reason about all three dimensions, as they need to coordinate considerations arising from the entire design space when generating responses. The \textit{Analysis Plan Critic}, focused on analytical strategies, addresses the \rhetoriccolor{rhetorical dimension}. Both the \textit{Code Critic} and the \textit{Visualization Critic} ensure the accurate execution of analytical strategies and validate the semantics of the results, thereby addressing both the \semanticcolor{semantic} and \rhetoriccolor{rhetorical} dimensions. The \textit{Interpretation \& Summary Critic} could potentially interpret the results, narrate strategies used, and provide actionable insights, encompassing all three dimensions. Thus, every dimension is covered by at least three agents in the system. \hl{Such extensive coverage ensures that responses are scrutinized multiple times, thereby reducing the likelihood of errors or overlooked considerations.}

We adopt a similar multi-agent framework for generating data stories (Figure~\ref{fig:story arch}). Given the Notebook and some user instructions, an \textit{Initial Respondent} generates the first draft. Three \textit{Critics}, each specializing in one dimension of the design space, then provide critiques based on their respective focus areas. (Since data stories are largely homogeneous in content type---natural language---we assign one \textit{Critic} to each dimension.) Following this, the \textit{Refiner} collaborates with the \textit{Critics} to improve the draft, incorporating their feedback, and then produces the final response.

\begin{figure*}[t]
  \centering
  \includegraphics[width=\textwidth]{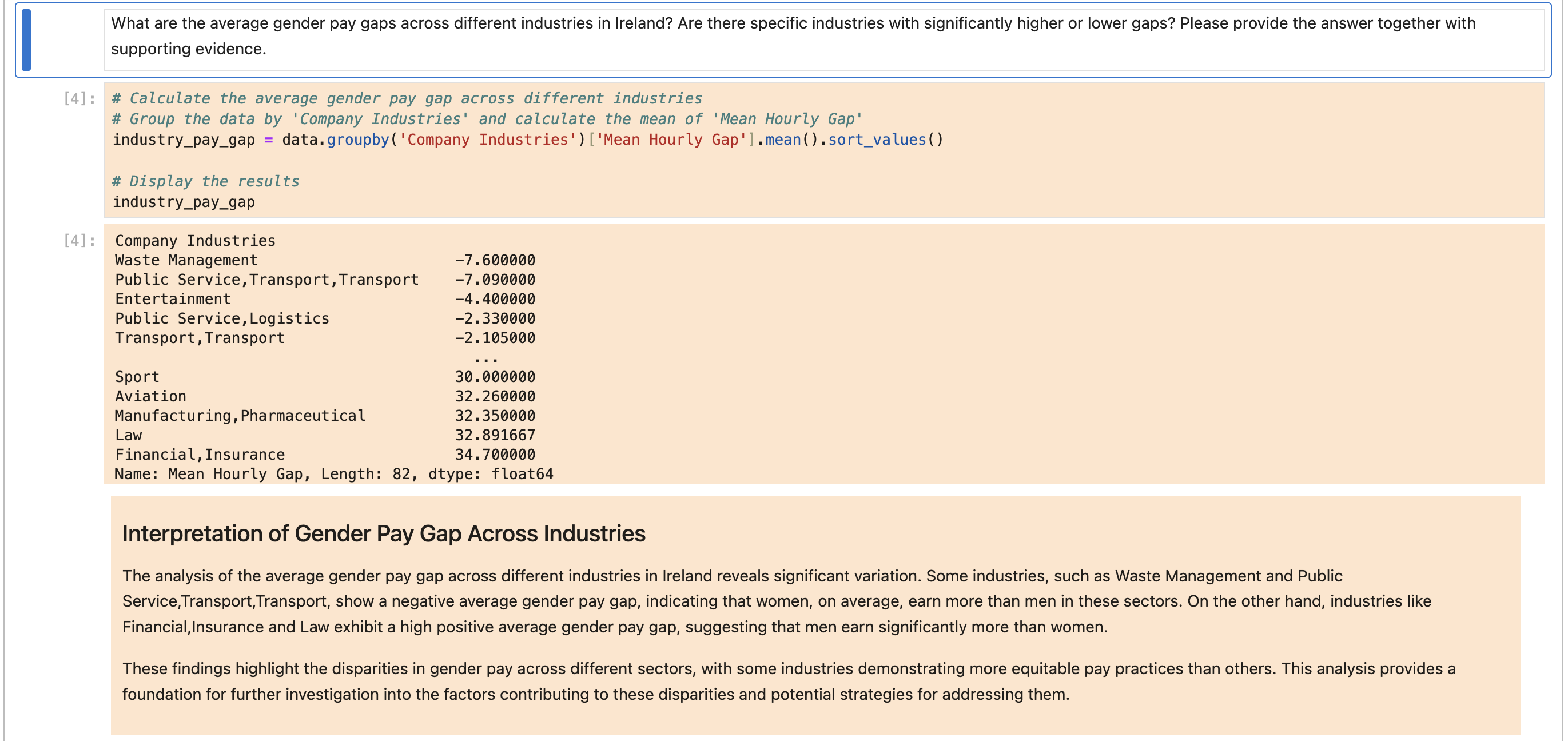}
  \caption{Rae tasks \sys\ with a complex query under the single-agent mode. Utilizing its agentic workflow, \sys\ produces code, executes it, and provides an interpretation. Nonetheless, this response is not immediately intuitive to Rae due to the lack of units, the limited coverage of only 10 industries, and a textual format that is difficult to parse. (The side panel is hidden for clearer presentation.)}
  \label{fig:eda single agent}
\end{figure*}

\subsection{Usage Scenario}
Rae, a data analyst at an NGO for gender equality, wants to analyze a dataset on the gender pay gap in Ireland~\cite{irelandData} and write a report with actionable insights. The dataset contains 1,178 records, detailing the pay disparities between male and female employees across various companies in Ireland. After installing \sys\ as a Jupyter Lab extension, she launches a new Notebook and starts her analysis. 

\begin{figure*}[t]
  \centering
  \includegraphics[width=\textwidth]{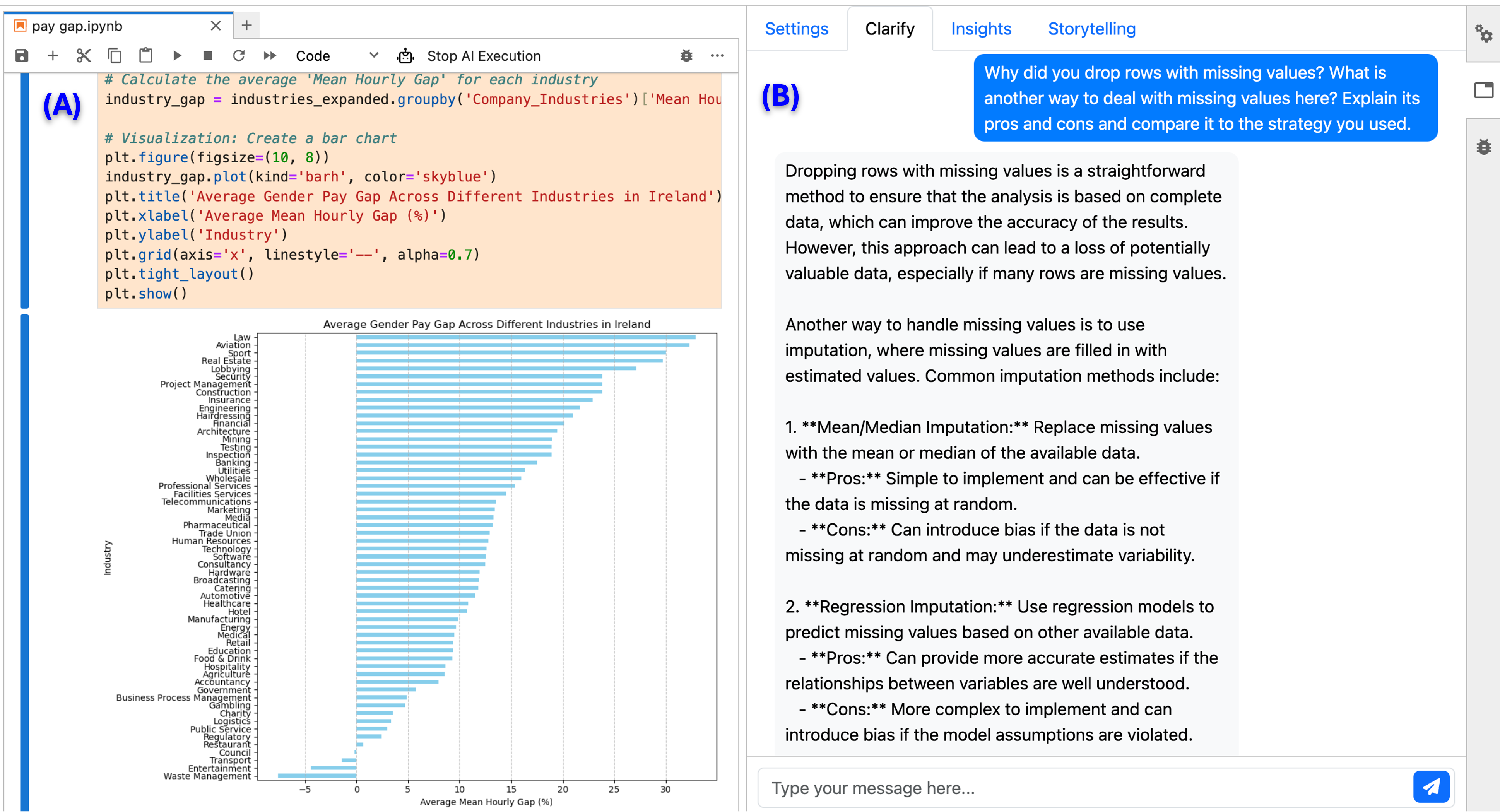}
  \caption{(A) Rae switches to the multi-agent mode for the previous complex query. (See the video walkthrough in the supplemental materials for how to configure this in the \textit{Settings} tab.) This time, \sys\ begins with an analysis plan in a markdown cell, executes the plan, produces a visualization, and finally generates an interpretation. Due to space constraints, only the code and the visualization are shown. Rae much prefers this response over the one in Figure~\ref{fig:eda single agent}. (B) Rae wants to understand the rationale for \sys's treatment of null values and alternative strategies. She engages in a threaded conversation with \sys\  in the \textit{Clarify} tab of the side panel.}
  \label{fig:clarify}
\end{figure*}

\subsubsection{EDA Copilot} Rae starts by creating a new cell in the Notebook and inputs her command to load the dataset~(Figure~\ref{fig:load dataset}). She sets \sys\ to the single-agent mode and clicks the ``robot'' icon in the cell toolbar (or in the Notebook toolbar) to invoke the AI. A loading icon appears in the Notebook toolbar, indicating that \sys\ is processing the request. After a few seconds, \sys\ inserts the response in a new cell below and executes it. The background of the LLM-generated cell is light peach-colored, allowing easy distinction from user-created cells. Concluding that the user request is fulfilled, \sys\ decides not to follow up, and the loading icon disappears.

After using \sys\ to get an overview of the dataset, Rae wants to gauge the average gender pay gaps across different industries in Ireland and identify those at the extremes. Still in single-agent mode, \sys\ first outputs the top and bottom five industries in terms of gender pay gap and then provides an interpretation of the result (Figure~\ref{fig:eda single agent}). However, the results are not immediately intuitive due to the lack of units, the limited coverage of only 10 industries, and a textual format that is difficult to parse. She then opens the \textit{Settings} tab in the side panel, toggles on the multi-agent mode for EDA, using GPT-4o for all agents and setting a max discussion round of one for the \textit{Critics} and the \textit{Refiner}, and reattempts this question (Figure~\ref{fig:clarify}(A)). This time, \sys\ begins with an analysis plan in a markdown cell, even accounting for steps such as handling missing values, then executes the plan, producing a clearly labeled visualization showing disparities across all industries in the dataset, and finally generates an interpretation in another markdown cell. Rae prefers this response and continues using the multi-agent mode for subsequent analyses.

\begin{figure*}[t]
  \centering
  \includegraphics[width=\textwidth]{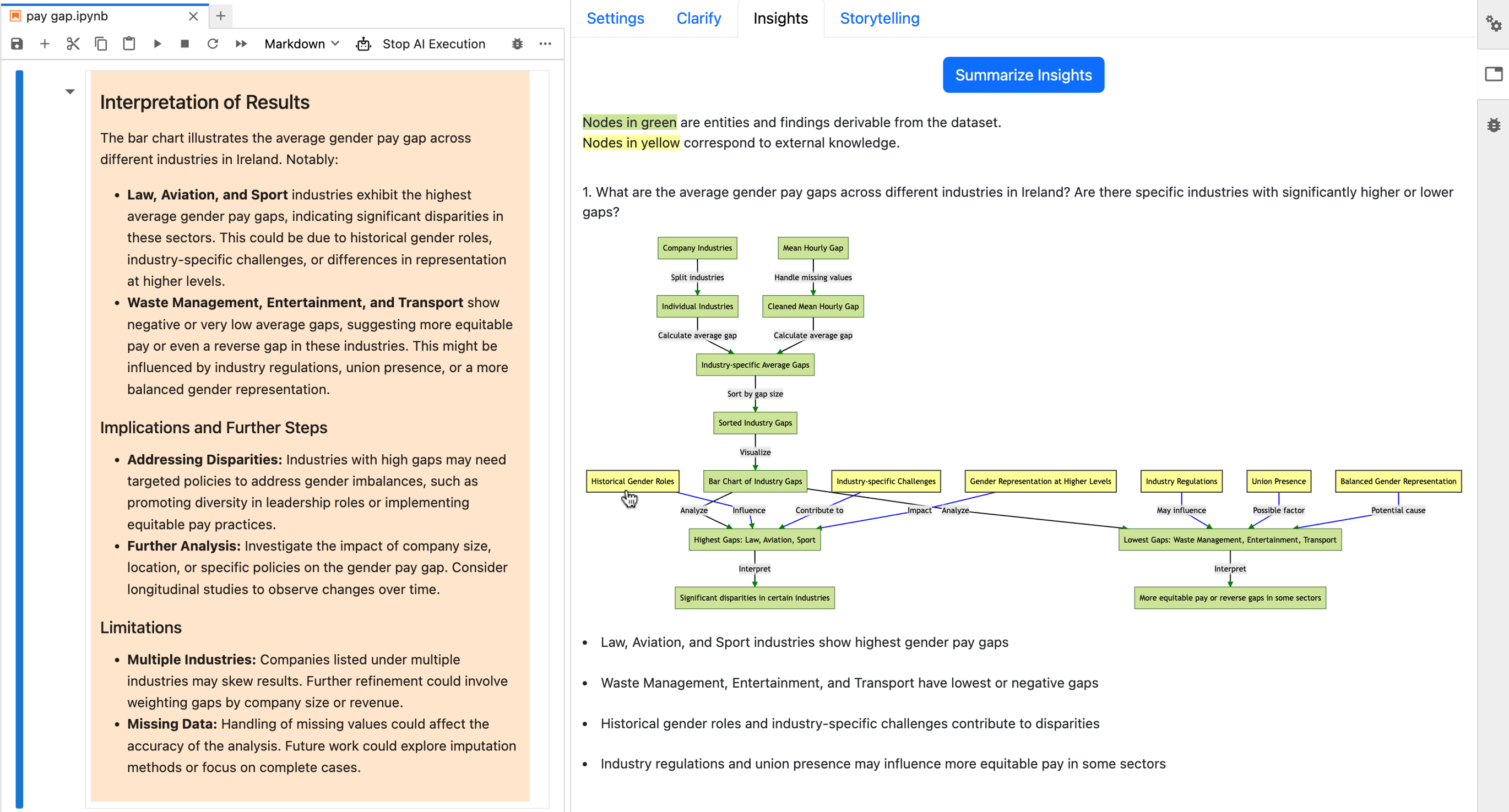}
  \caption{Rae wants to summarize the analysis history and insights gained so far. She navigates to the \textit{Insights} tab and selects ``Summarize Insights.'' \sys\ generates a graphical representation of the insights for each analytical question explored in the Notebook. Rae can also click on any of the nodes or edges in the graph, and \sys\ will scroll to and highlight the most relevant cell in the Notebook.}
  \label{fig:insights}
\end{figure*}

\subsubsection{Clarification} Perusing the LLM-generated code for the previous response (Figure~\ref{fig:clarify}(A)), Rae is curious why missing values were dropped and wonders about alternative strategies for handling them. Not wanting to disrupt the flow of the Notebook, she opts to take this clarification to a separate interface. She opens the \textit{Clarification} tab in the side panel, selects the Notebook cell in question, and inputs her query (Figure~\ref{fig:clarify}(B)). From the LLM's response, Rae learns that dropping records with null values is the simplest strategy and considers whether to explore other options.

\subsubsection{Tracking Insights} Rae now wants to review the analysis history along with insights she gained so far. She navigates to the \textit{Insights} tab and clicks ``Summarize Insights'' (Figure~\ref{fig:insights}). \sys\ then summarizes the Notebook, organized by each analytical question. Rae is able to quickly trace the the graphical representation of the analysis history and refresh her memory on key findings, such as the industries with the highest pay gaps, as well as the data transformations that led to these results.

Reading from the graph that historical gender roles might have contributed to high gender pay gaps in law, aviation, and sports, Rae wants to locate the cell in the Notebook where this was mentioned to revisit it. She clicks the node in question, and \sys\ scrolls to and highlights the target cell in the Notebook. Reviewing the summary helps Rae marshal the findings. She is now better prepared for further analysis.

\begin{figure*}[t]
  \centering
  \includegraphics[width=\textwidth]{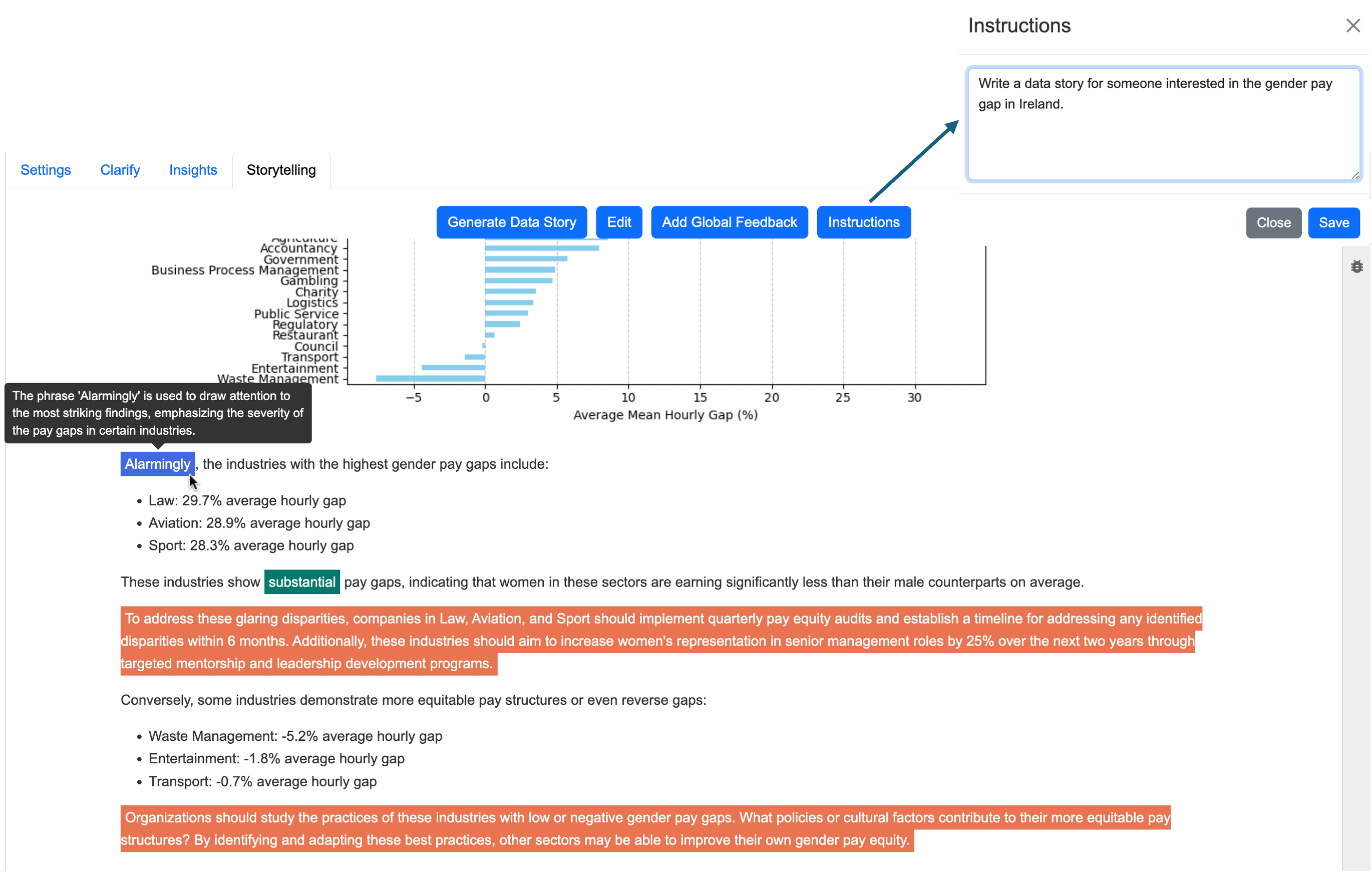}
  \caption{Rae navigates to the \textit{Settings} tab to activate the multi-agent mode for data storytelling (not shown here; please see video supplement). To generate a data story, Rae navigates to the \textit{Storytelling} tab. She inputs her instructions and clicks ``Generate Data Story.'' \sys\ produces a coherent and persuasive data story as an HTML page, highlighting sections corresponding to the three dimensions, with explanations provided in tooltips. (The left panel is hidden for clearer presentation.)}
  \label{fig:generate data story}
\end{figure*}

\subsubsection{Generating a Data Story}
Rae discovers some interesting results and now wants to generate a data story. She opens the \textit{Settings} tab, enables the multi-agent mode for data storytelling, sets all agents to Claude~3.5~Sonnet, and limits the discussion rounds between the \textit{Critics} and the \textit{Refiner} to one. Next, she navigates to the \textit{Storytelling} tab, clicks ``Instructions'' to input her request to ``[w]rite a data story for someone interested in the gender pay gap'' in a modal box (Figure~\ref{fig:generate data story}), and then clicks the ``Generate Data Story'' button. In response, \sys\ returns an HTML page containing the data story. Scrolling through the data story, Rae is pleasantly surprised by its quality, as the story is grounded in data facts (\eg citing statistics from the analysis highlighting the range of gender pay gaps), uses clear and persuasive language (\eg the word ``substantial'' accurately captures the severity of the pay gaps), and recommends sensible courses of actions (\eg implementing quarterly pay equity audits). 

Rae sees some text is highlighted and hovers over it. Tooltips appear, explaining the language choices or the bases for the insights. For example, Rae learns that the word ``alarmingly'' serves a \rhetoriccolor{rhetorical} purpose by drawing the reader's attention to the most striking findings and underscoring the severity of the pay gaps in certain industries. These explanations not only provide transparency to Rae, but also offer her entry points for editing the data story when her thoughts do not align with the current narrative.

\begin{figure*}[t]
  \centering
  \includegraphics[width=\textwidth]{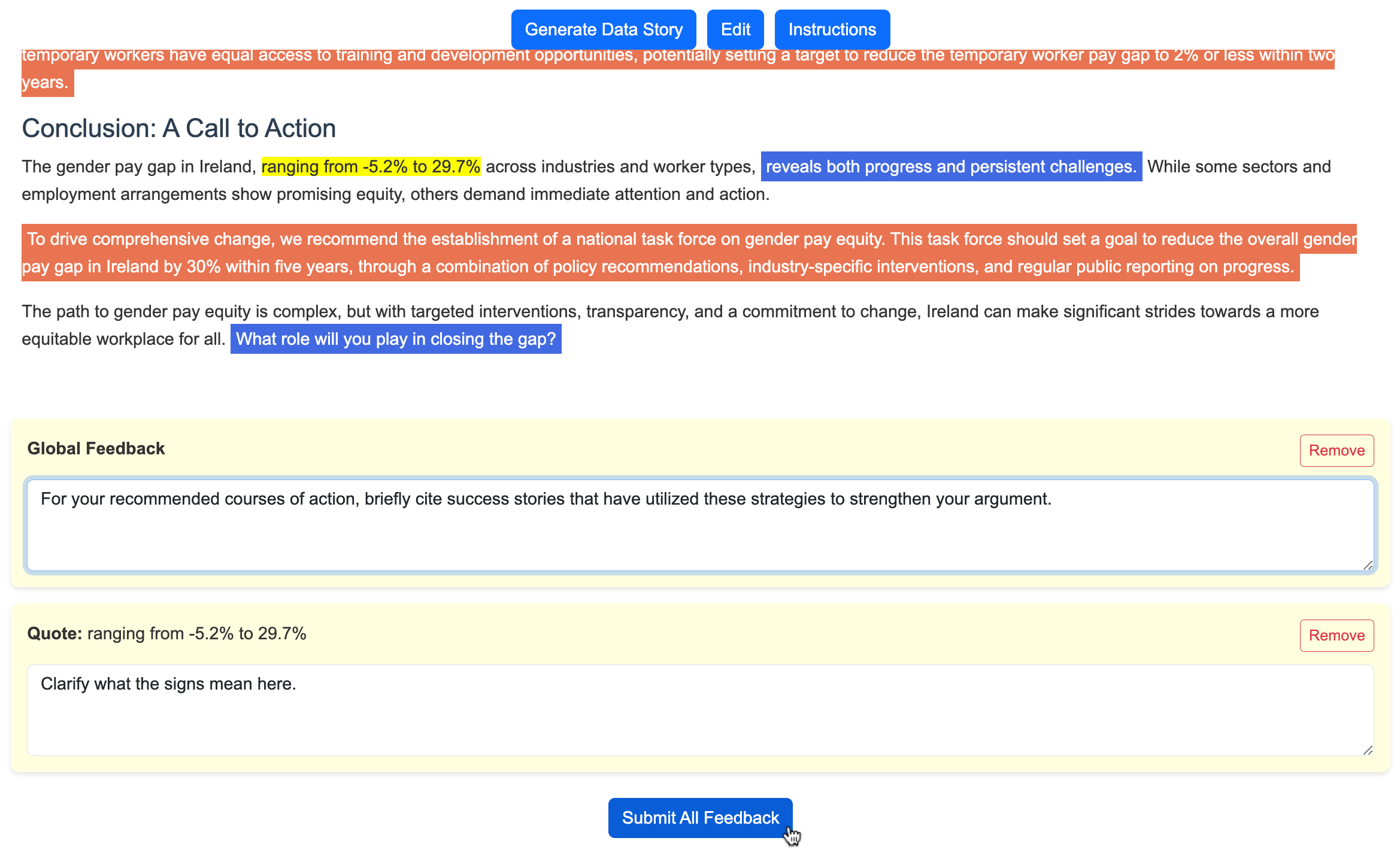}
  \caption{Rae can provide both global feedback on the entire AI-generated data story and local feedback focused on specific parts. After clicking ``Submit All Feedback,'' \sys\ refines the story according to her requests. (The left panel is hidden for clearer presentation.)}
  \label{fig:ai edit}
\end{figure*}

\begin{figure*}[t]
  \centering
  \includegraphics[width=\textwidth]{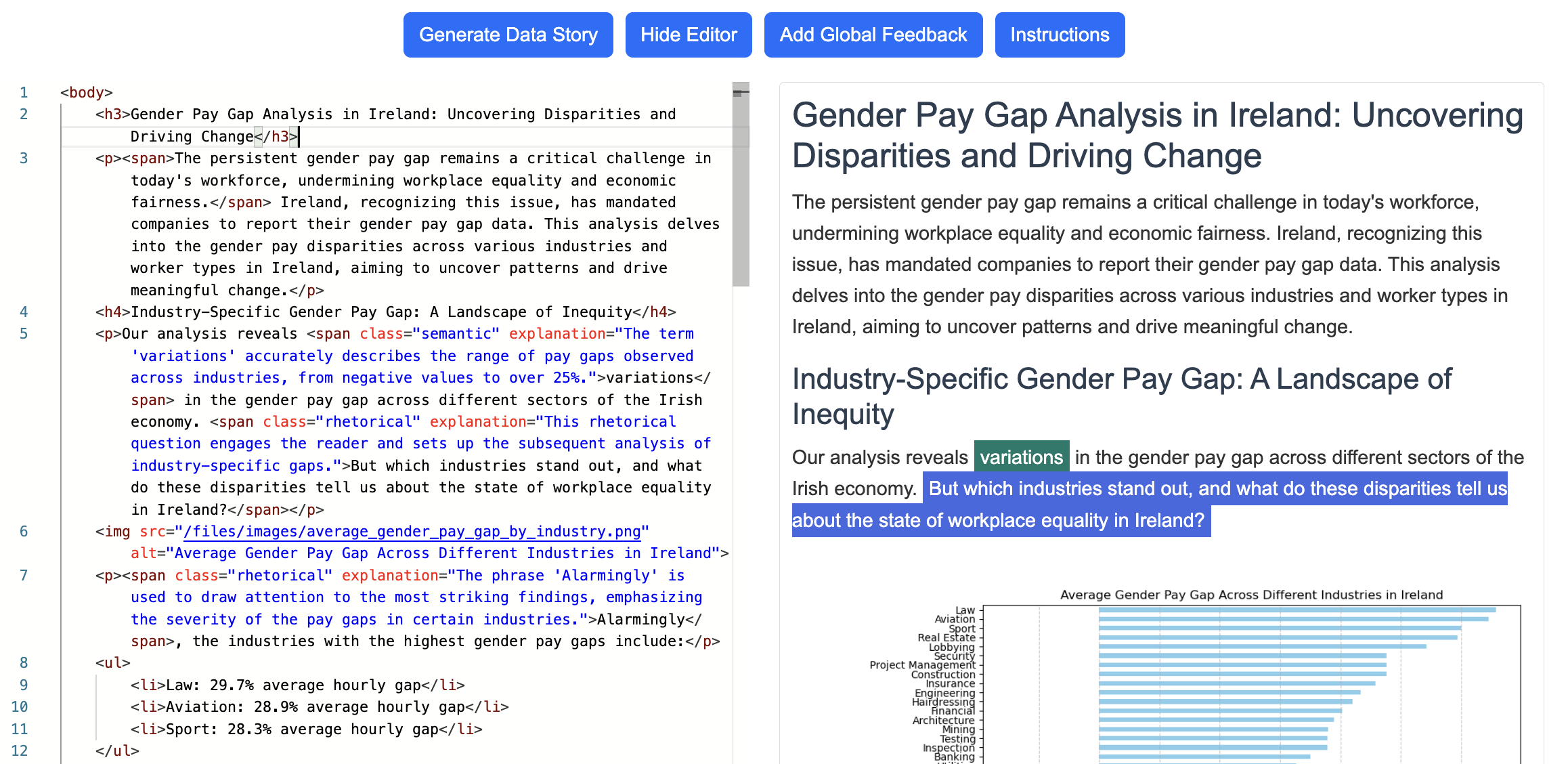}
  \caption{Rae can also manually edit the data story by clicking ``Edit,'' which opens a side-by-side live HTML editor next to the rendered data story. (The left panel is hidden for clearer presentation.)}
  \label{fig:editor}
\end{figure*}

\subsubsection{Editing the Data Story}
While Rae generally likes the data story, she wants to fine-tune it. For instance, she hopes to strengthen the narrative by adding stories that illustrate how the strategies recommended by \sys\ have been successfully used to combat the gender pay gap. She enlists AI assistance by clicking ``Add Global Feedback'' and inputs her request as a piece of global feedback in a (Figure~\ref{fig:ai edit}). She also wants the data story to be clear about the sign (plus or minus) of the pay gap, so she selects the relevant section, clicks a popup button to ``Add a Note,'' and instructs \sys\ accordingly. After clicking ``Submit All Feedback,'' the AI updates the data story momentarily.

Finally, Rae decides to manually edit the data story. She clicks the ``Edit'' button, which opens a side-by-side live HTML editor next to the rendered page (Figure~\ref{fig:editor}). After making her final adjustments, she exports the content as an HTML web page, ready to circulate the report on social media to raise awareness.

\section{Evaluation}

To evaluate \sys\ and compare its two modes (\ie single- vs. multi-agent), we conducted a user study with nine experienced data analysts. In our study, each participant analyzed a dataset in their area of expertise and crafted a data story using \sys\ with the goal of conveying actionable insights. Through questionnaires and semi-structured interviews, we assessed \sys's strengths and limitations.

\subsection{Participants}
\hl{We recruited participants from X (formerly Twitter), LinkedIn, and three Slack workspaces frequented by data analysts.} Nine expert data analysts (\textbf{E1} - \textbf{E9}\hl{; five men and four women}) with extensive experience in (1) actionable EDA and data storytelling and (2) using AI tools for data analysis \hl{participated in our study, including two who also took part in our formative interview}. Three were PhD students in data-science-adjacent disciplines, while the rest were professional data scientists, consultants, and machine learning engineers. Participants reported an average of 5.9 years of data analysis experience. They also brought diverse domain expertise to the task, including in finance, retail, and healthcare. \hl{As with the formative interview, we screened participants through an intake survey focusing on their experience in actionable EDA and storytelling; all participants also demonstrated proficiency in data analysis during the sessions.} We rewarded each participant with a \$30 gift card for an approximately 65-minute session.

\subsection{Task}
\label{task}
We instructed participants to analyze a dataset and create a data story to convey actionable insights using \sys. To minimize the risk of LLMs reproducing training data, we encouraged participants to bring a \textbf{private, non-synthetic dataset} they had previously worked on. Six participants provided their own datasets. For each of the remaining three without a private dataset, we found an online dataset tailored to their expertise that had been released after the knowledge cutoff of the LLMs used (GPT-4o\footnote{The knowledge cutoff of GPT-4o is October 2023~\cite{gpt4o}.} and Claude~3.5~Sonnet\footnote{The knowledge cutoff of Claude~3.5~Sonnet is April 2024~\cite{claude}.}). We now detail our study protocol:

\paragraph{Introduction and System Walkthrough (\textasciitilde{}10 minutes)} We first walked participants through the system interface and played a video demonstration of \sys\ while narrating its key features. Participants then loaded their dataset using \sys\ to commence the task.

\paragraph{EDA (\textasciitilde{}30 minutes)} Participants were instructed to freely explore the dataset and extract actionable insights, starting with the single-agent mode of \sys. Then, they repeated the same analysis using the multi-agent mode. We encouraged participants to utilize the clarification and insight summary features as needed. We also explicitly asked participants to take a measured approach, thoroughly reviewing the analysis plans, code, visualizations, and interpretations while thinking out loud.

\paragraph{Data Storytelling (\textasciitilde{}15 minutes)} Participants provided instructions and generated two data stories using \sys, first with the single-agent mode and then with the multi-agent mode. They perused the data stories, along with the LLM-generated explanations in the tooltips, while thinking out loud. Participants then refined the second data story via user-guided AI refinement. Finally, they finetuned the data story using the live HTML editor.

\paragraph{Questionnaire and Semi-Structured Interview (\textasciitilde{}10 minutes)} Participants completed a questionnaire on the usefulness and usability of \sys\ using a 5-point Likert scale. The questionnaire comprised two parts. The first part asked participants to separately rate their experience using the data analysis plugin in ChatGPT and \sys\ for assisting with actionable EDA and storytelling. While participants did not use the data analysis feature in ChatGPT during the study due to time constraints, all had prior experience with it; in fact, five participants had previously analyzed the dataset they brought using ChatGPT. Specifically, we assessed how ``enjoyable,'' ``usable,'' ``helpful,'' ``integrated into your workflow'' (\textbf{D1}), ``steerable'' (\textbf{D3}), ``explainable'' (\textbf{D4}), and ``reparable'' (\textbf{D5}) the systems were. The second part of the questionnaire focused on evaluating the quality of responses across the three dimensions of our design space---the extent to which the systems ``provides precise and contextually rich interpretation of results,'' ``generates coherent and persuasive analyses and narratives,'' and ``offers high-quality actionable insights''---for the single- and multi-agent modes of \sys. We then followed up with a semi-structured interview to gain deeper insights into user experience and gather suggestions for future improvements.

\begin{figure*}[t]
  \centering
  \includegraphics[width=\textwidth]{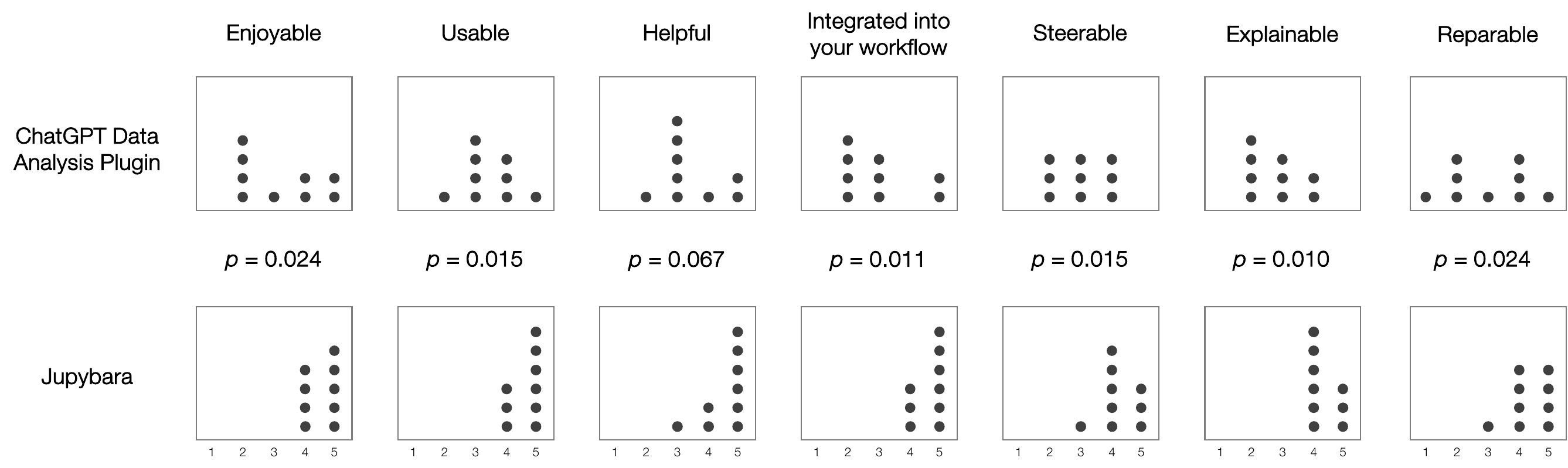}
  \caption{Participants separately rated ChatGPT's data analysis plugin and \sys\ on how ``enjoyable,'' ``usable,'' ``helpful,'' ``integrated into [their] workflow,'' ``steerable,'' ``explainable,'' and ``reparable'' they were for assisting with actionable EDA and storytelling. \sys\ achieved higher median ratings across all dimensions. A Wilcoxon signed-rank test was conducted to compare the two systems on each dimension. After applying the Holm-Bonferroni correction, the \textit{p} values do not indicate a significant difference, likely due to the small sample size, but there is a clear trend that participants preferred \sys\ across all dimensions.}
  \label{fig:gpt vs jupybara}
\end{figure*}

\begin{figure*}[t]
  \centering
  \includegraphics[width=0.56\textwidth]{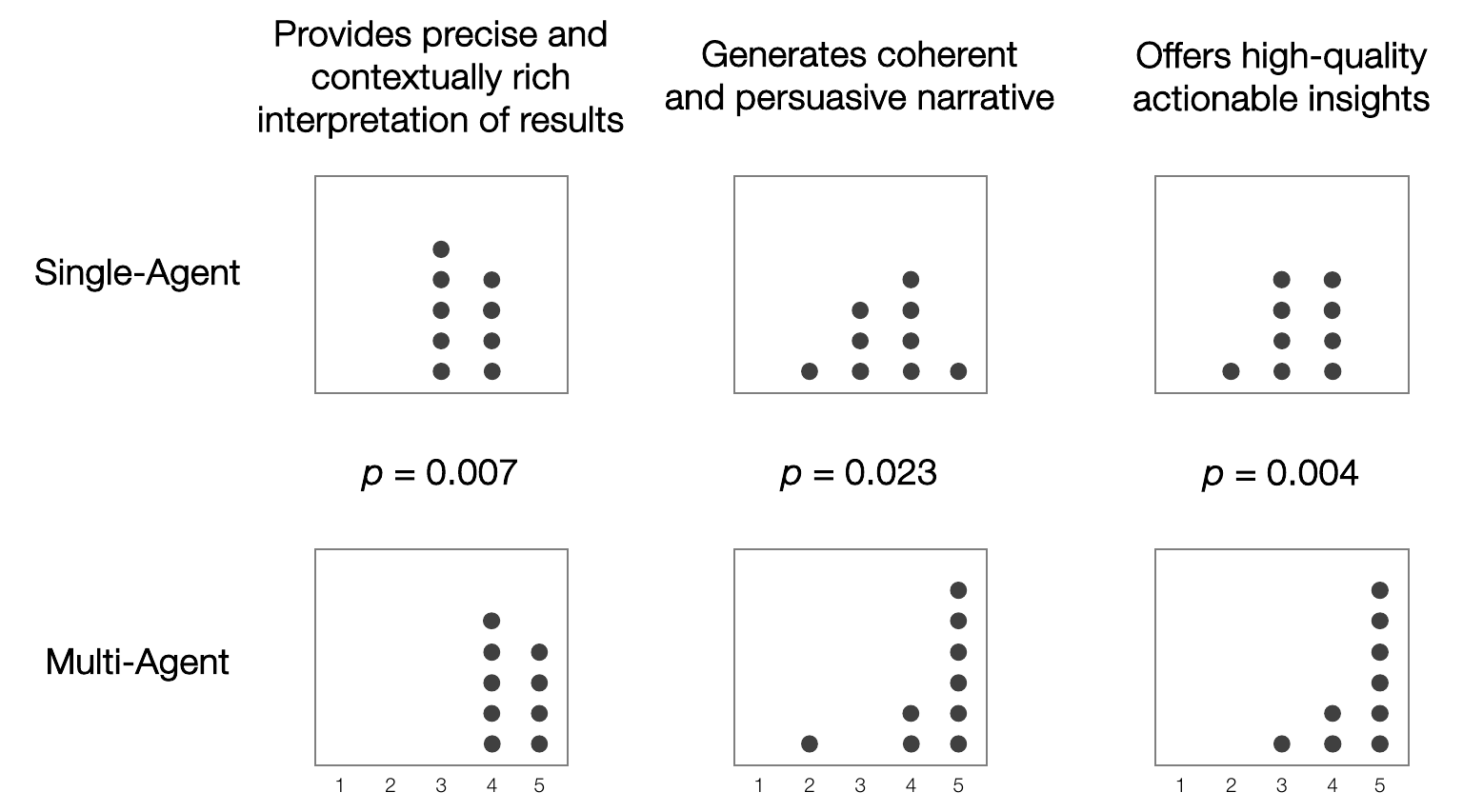}
  \caption{Participants separately rated the single- and multi-agent modes of \sys\ on the three dimensions of our design space. For every dimension, the multi-agent mode achieved a higher median rating, scoring either 4 or 5. A Wilcoxon signed-rank test comparing the two modes also indicates a significant difference in favor of the multi-agent mode for each dimension, even after the Holm-Bonferroni correction, suggesting participants generally preferred the responses generated by the multi-agent mode.}
  \label{fig:single vs multi}
\end{figure*}

\subsection{Results}
\subsubsection{Questionnaire Results} Figure~\ref{fig:gpt vs jupybara} presents participants' ratings of ChatGPT's data analysis plugin and \sys\ on seven measures for supporting actionable EDA and storytelling. Compared to ChatGPT's data analysis plugin, \sys\ achieved higher median ratings across all dimensions, with values of either 4 or 5. We also conducted a Wilcoxon signed-rank test~\cite{wilcoxon1992individual} comparing the two systems on each dimension. Although the \textit{p} values do not indicate significant differences after applying the Holm-Bonferroni correction~\cite{holm1979simple} (Figure~\ref{fig:gpt vs jupybara}), likely due to the small sample size, the trend clearly indicates that participants rated \sys\ as superior across all dimensions.

Figure~\ref{fig:single vs multi} presents participants' ratings of the single- and multi-agent modes of \sys\ on the three dimensions of our design space. For every dimension, the multi-agent mode achieved a higher median rating, scoring either 4 or 5. A Wilcoxon signed-rank test comparing the two modes also indicated a significant difference in favor of the multi-agent mode for each dimension even after the Holm-Bonferroni correction (Figure~\ref{fig:single vs multi}), suggesting participants generally preferred the responses generated by the multi-agent mode.

\subsubsection{General Usability}
Participants widely recognized the high usability of \sys. In particular, participants appreciated \sys's integration into their workflows as a Jupyter Notebook extension. They were able to ``\textit{do everything in Jupyter}'' and did not need to ``\textit{copy and paste between Jupyter and ChatGPT}'' all the time, which was especially cumbersome when dealing with visualizations and error messages. In our session with \textbf{E6}, he was pleasantly surprised when \sys\ installed a geo-visualization package using command line input in Jupyter Notebook (with the ``\verb|!|'' shell escape) and created an interactive map visualizing the distribution of restaurants in a city.

Participants also praised the intuitive design of the interface and found it easy to interact with the AI. As \textbf{E8} said, ``\textit{I like how easy it is to tell which cells are AI-generated (because of the differences in background color).}'' All participants quickly learned to interact with the EDA copilot, likely because the paradigm of asking questions in one cell and receiving AI responses in the cells below resembles the familiar turn-taking chat experience of ChatGPT. The side panel further enhanced participants' ability to cross-reference the Notebook with data stories, summaries, or threaded conversations. The tabbed design within the side panel also received positive feedback for clearly separating different functionalities.

\subsubsection{Usage Patterns and Usefulness of Features}
We observed participants' workflows when using \sys\ and synthesized their perspectives on the usefulness of each feature. Overall, participants spent the most time with the EDA copilot and the data story generation and revision features, which they also regarded as the most useful functionalities of the system. While this may be partly due to the way the user study was structured (Section~\ref{task}), it is not surprising, as these features provide direct, automated assistance for EDA and storytelling. Most participants expressed enthusiasm for the clarification feature. Interestingly, we observed varied usage patterns: while some used the feature mainly for answering questions about analytical strategies (\eg how to deal with multi-collinearity), others used it to gain more context about the actionable insights or even to challenge the system’s interpretations and recommendations. Regarding the insights tracking feature, participants praised the graphical representation for its clarity and conciseness, preferring that over the ``\textit{wall of text}'' (\textbf{E3}) that ChatGPT typically produces. While some participants found it unnecessary to generate such summaries for their analyses, which were often limited in scope and depth due to the time constraint of our study, most agreed that this feature would be beneficial for summarizing longer notebooks or even for presentation purposes.

\subsubsection{Steerability}
During EDA, participants utilized \sys\ to perform a wide variety of tasks, including data cleaning, visualization, and model building. We also observed a mix of close-ended commands (e.g., renaming columns) and open-ended commands (e.g., suggesting and exploring complex problems). In general, participants felt ``\textit{unconstrained in what instructions they could give}'' (\textbf{E2}) and that \sys\ was capable of following both concrete and more abstract, or even ambiguous, instructions. Most participants (\textit{n} = 7) also believed that \sys\ more effectively interpreted and executed their instructions than the web-based ChatGPT. We attribute this to our use of the ReACT prompting paradigm and our strategies for operationalizing the design space. In addition, participants deemed \sys's agentic behavior---the ability to create new cells, execute them, and automatically follow up---as invaluable to handling complex tasks. In data storytelling, participants appreciated the ability to guide how the data story should be generated through the modal input, which \sys\ mostly adhered to. For instance, \textbf{E8}, who experimented with generating data stories answering different analytical questions and for different audiences, was intrigued by \sys's ability to ``\textit{adapt its response to [their] request.}''

\subsubsection{Explainability}
Participants generally reported experiencing more transparency with \sys, especially in the multi-agent mode, than with ChatGPT for actionable EDA and storytelling. \textbf{E4} noted, ``\textit{\sys\ often gives me a heads-up on how to analyze the data before writing code, which helps me better understand its tactics.}'' \textbf{E7} commented on the explainability exhibited in the data stories: ``\textit{The tooltips are pretty useful, especially the ones for actionable insights. I can use this information to judge if they are valid insights.}'' Additionally, participants welcomed the LLM-generated code comments. \textbf{E4} said, ``\textit{I noticed \sys\ added comments about possible outcomes from the code and gives interpretations for each of them.}'' This feature helps users stay informed about multiple potential outcomes, not just the one they found, offering a broader understanding of their analysis. These observations highlight the success of our operationalization strategies. By offering a framework in which LLMs can refine their responses, \sys\ also enables the LLMs to explain themselves more effectively along these dimensions. This combination of proactive explanations (code comments and tooltips) and user-driven clarification (the \textit{Clarification} tab) contributes to a more transparent user experience. That said, two participants commented that \sys\ highlighted too much text for explanation in the tooltips, which could be overwhelming. Additionally, some of the explanations, such as ```significantly increases' accurately describes the trend,'' were not very informative. These observations suggest room for improvement in providing essential and helpful explanations.

\subsubsection{Reparability}
All participants agreed that \sys\ demonstrated high reparability. On one hand, its agentic characteristics often allowed the system to self-repair in carrying out EDA. In the presence of bugs, \sys\ often managed to correct and recover based on error messages. Perhaps more notably, due to the emphasis on valid semantics in operationalizing the design space, \sys\ even proactively fixed some non-error issues. As \textbf{E6} remarked, ``\textit{I'm amazed by how \sys\ automatically remakes visualizations when there’s clutter in the visualizations. These are not errors, but it still fixes them. ChatGPT would only fix errors.}''

On the other hand, participants appreciated having the option to initiate repairs on AI-generated content. \textbf{E2} said, ``\textit{I think it's great that I can always edit AI code and data reports.}'' The explanations provided by the AI further facilitated the editing process. As \textbf{E2} also noted, ``\textit{when I don't agree with what the AI says, I have a better sense of how to fix it.}'' The ability to ask AI to fix their responses---by appending new commands in EDA or providing feedback in data storytelling---was also popular among participants. In fact, we observed that participants generally preferred requesting AI adjustment as a first step before making manual refinement, and oftentimes the AI was capable of addressing these requests.

\subsubsection{Single- vs. Multi-Agent Modes}
All but one participant preferred the multi-agent mode over the single-agent mode in terms of the quality of responses. In EDA, participants noted that the multi-agent mode produced more robust plans and made the analysis more digestible through clear visualizations and explanations. This mode also tended to be more detail-oriented (\eg checking conditions for statistical tests such as normality), and was additionally more resourceful. For instance, the multi-agent mode not only utilized statistical machine learning models but also suggested and implemented neural networks for \textbf{E5}'s dataset.

When writing data stories, the multi-agent mode provided more contextually rich and accurate descriptions of results (\eg describing a basketball player who scored high on multiple metrics as a ``versatile player''). Across a wide range of domains, the multi-agent mode produced high-quality actionable insights. For example, the insights suggested by \sys\ for the private survey responses \textbf{E9} collected closely matched what she had written in an unpublished manuscript on computer science education, showcasing \sys's ability to apply domain expertise to data facts. Moreover, the multi-agent mode more effectively and reliably cited external sources to support actionable insights, such as historical events or academic publications, which, upon verification, proved accurate.

Despite the clear advantages in response quality, other factors also influenced participants' preferences on which mode to use.  One such factor is latency. Since the multi-agent mode involves multiple queries to LLMs, its response time is approximately five times longer than that of the single-agent mode when the maximum discussion rounds between the \textit{Critics} and the \textit{Refiner} are set to two, which can negatively affect user experience. Additionally, while multi-agent EDA responses tend to be comprehensive, they can also be more verbose (\textbf{E7}). Consequently, when asked about their potential usage of \sys, many participants indicated they would dynamically adjust settings based on task complexity, opting for the single-agent mode for straightforward queries and the multi-agent mode for more intricate analyses.

\section{Limitations}
\subsection{User Study Limitation}
\hl{Our user studies were limited by relatively small participant pools, with nine participants in both the formative and summative studies. The requirement for extensive experience in actionable EDA and storytelling led us to exclude many respondents from our intake survey, such as undergraduate students with limited experience analyzing real-world datasets and contributing to decision-making processes. Furthermore, the 65-minute summative evaluation may have been insufficient for participants to thoroughly explore \textsc{Jupybara}\ or assess its performance on complex, real-world tasks. To address these limitations, we plan to open-source \textsc{Jupybara}\ and promote its integration into daily data analysis workflows, particularly for large-scale, long-term projects, to gather more comprehensive feedback and further refine the system.}

\hl{Another limitation in our summative evaluation was that participants did not use ChatGPT's data analysis plugin during the study due to time constraints. Nonetheless, all participants had prior experience with the ChatGPT plugin, and five had used it to analyze the datasets they provided. Without a direct comparison of both systems in the study, we caution that participants’ higher ratings of \textsc{Jupybara}\ may have been partially influenced by \textsc{Jupybara}'s novelty, which could have positively biased their preferences. Furthermore, participants’ preference for the multi-agent mode over the single-agent mode may have been influenced by the order of use, as they always started with the single-agent mode before transitioning to the multi-agent mode, which potentially introduced a learning bias as they became more familiar with the system. Additionally, the longer response times in the multi-agent mode might have led participants to perceive its outputs as more thorough, conflating processing time with output quality.}

\subsection{System Limitation}
\hl{During the study, we observed several limitations of \textsc{Jupybara}. While \textsc{Jupybara}\ was generally effective at fixing execution errors, there were three instances where the system failed to self-correct and was trapped in a cycle of recurring errors without considering alternative approaches. An illustrative example was when \textsc{Jupybara}\ consistently mistook the data type of a column in a dataframe. In such cases, users had to manually intervene or switch to another language model. As an LLM-powered system, \textsc{Jupybara}'s coding ability is constrained by the underlying language model, but a better agent architecture, coupled with more graceful exit strategies, might better help users handle such situations.}

\hl{Another limitation concerns system responsiveness. Unlike major web-based chat interfaces that provide streaming responses from LLMs, \textsc{Jupybara}\ delivers responses all at once for each cell. Our approach involves parsing LLM-generated JSON objects to extract metadata (e.g., cell type) when creating new cells. Since parsing can only occur after the entire JSON is generated, it may occasionally result in perceived delays. Some users raised concerns about the system's responsiveness, especially when handling complex tasks that require more processing time.}

\hl{Moreover, the multi-agent architecture, while enhancing the quality of responses, introduces latency due to the additional computational overhead of agent interactions. Two participants expressed a desire for more detailed progress updates beyond the standard ``loading'' icon to alleviate concerns about system inactivity. %The internal thought processes of the LLM agents are not sufficiently surfaced in the multi-agent mode, leaving users feeling as though they are waiting without feedback. 
Providing more granular progress indicators or exposing some of the agents' intermediate reasoning could improve the user experience by making the system's operations more transparent during these periods.
}

\section{Future Work}

\sys\ represents our first effort to design an LLM-based intelligent agent to assist data analysts in actionable EDA and storytelling based on the design space framework we have developed. Future work should focus on improving the interface and interaction design of \sys\ and introducing new, helpful functionalities. \hl{For example, beyond outputting data stories as documents, \textsc{Jupybara} could be enhanced to support other formats, such as presentation slides or data videos.} We also believe \sys\ would benefit from a feature that automatically chooses between the single- and multi-agent modes based on query complexity, balancing latency and response quality. \hl{In addition, future work could explore adopting a progressive view for tracking insights, where insights are initially summarized as succinct statements and can be expanded into the graphical layout on demand, boosting system responsiveness and usability.}
%our evaluation revealed that data analysts would appreciate better communication of LLM ``thought processes'' to make more informed decisions when collaborating with the AI

While our work demonstrates the effectiveness of design-space-aware prompting and multi-agent architectures, alternative prompting strategies and agent configurations may well prove superior. We also believe there are additional techniques worth exploring, such as incorporating retrieval-augmented generation (RAG) into the pipeline. %\vidya{why?}. 
This is motivated by the fact that different domains tend to observe distinct analytic and linguistic conventions in actionable EDA and storytelling, which general-purpose LLMs may not fully grasp. With a carefully curated, domain-specific repository of best practices and sample analyses, \sys\ could leverage RAG to provide even higher-quality responses. We encourage future research to refine and expand upon our strategies for operationalizing the design space for EDA and storytelling aimed at producing and conveying actionable insights. 

Finally, we believe \sys\ can serve as a rich testbed for exploring how data analysts collaborate with AI and for developing design guidelines for the creation of AI-driven tools in data science. Integration into the popular Jupyter platform makes the system an ecologically valid design probe, and the system's multiple features and functionalities allow for the study of diverse interactions and workflows. For example, using \sys, researchers might examine the effects of different strategies for surfacing multi-agent interactions to data analysts on their trust and decision-making. \hl{Additionally, while \textsc{Jupybara} supports a high level of automation, users can also flexibly prompt the system to operate at varying levels of agency. For instance, \textsc{Jupybara} can be directed to outline a plan without executing it for user verification, or act as a copilot that reviews user-led analysis. This flexibility, driven by user prompts, allows for diverse usage modes, making it adaptable to the needs of data science learners, novices, and experts. Future work could explore how \textsc{Jupybara} can serve as a probe to study the optimal level of agency for different user groups and contexts.} These findings will, in turn, help shape better AI assistants for EDA and storytelling. Upon acceptance of our paper, we will open-source \sys\ and gather feedback from the user community to further enhance its usability and usefulness.

\section{Conclusion}

We contribute a comprehensive design space for actionable EDA and storytelling integrating the \semanticcolor{semantic}, \rhetoriccolor{rhetorical}, and \pragmaticcolor{pragmatic} dimensions, derived from foundational theories in data visualization, narrative discourse, and communication, and validated through expert interviews. By operationalizing this framework through design-space-aware prompting and multi-agent architectures, we develop \sys, an AI-enabled assistant implemented as a Jupyter Notebook extension. Our expert evaluation demonstrates that \sys\ effectively assists analysts in data exploration and conveying actionable insights. We encourage future work to leverage \sys\ as an ecologically valid testbed for exploring human-AI collaboration across a broader range of analytical tasks and settings.

\bibliographystyle{ACM-Reference-Format}
\bibliography{main}

\end{document}